# On the Cognitive Interference Channel with Unidirectional Destination Cooperation

Hsuan-Yi Chu and Hsuan-Jung Su

Graduate Institute of Communication Engineering

Department of Electrical Engineering

National Taiwan University, Taipei, Taiwan

Email: b95901210@ntu.edu.tw, hjsu@cc.ee.ntu.edu.tw

*Abstract*—The cognitive interference channel with unidirectional destination cooperation (CIFC-UDC) is a cognitive interference channel (CIFC) where the cognitive (secondary) destination not only decodes the information sent from its sending dual but also helps enhance the communication of the primary user. This channel model is an extension of the original CIFC to achieve a win-win solution under the coexistence condition. From an information-theoretic perspective, the CIFC-UDC comprises a broadcast channel (BC), a relay channel (RC) and a partially cooperative relay broadcast channel (PCRBC), and can be degraded to any one of them.

Our main result is the establishment of a new unified achievable rate region for the CIFC-UDC which is the largest known to date and can be explicitly shown to include the previous result proposed by Chu and the largest known rate regions for the BC, the RC and the PCRBC. In addition, an interesting viewpoint on the unidirectional destination cooperation in the CIFC-UDC is discussed: to enable the decoder of the primary user to perform interference mitigation can be considered as a complementary idea to the interference mitigation via Gel'fand-Pinsker precoding in the CIFC proposed by Devroye et al. Henceforth, by combing these two ideas, the interferences caused at both the destinations can be alleviated. Lastly, an outer bound is presented and proved to be tight for a class of the CIFC-UDC, resulting in the characterization of the capacity region for this class.

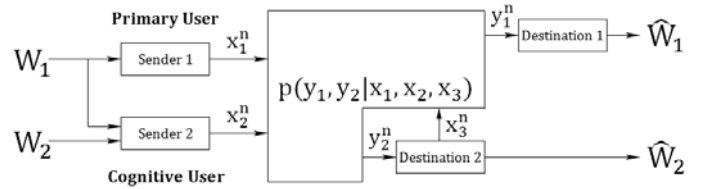

Fig. 1. The channel model of the cognitive interference channel with unidirectional destination cooperation (CIFC-UDC).

## I. INTRODUCTION

The rapid advancement of wireless technology has facilitated the development and growth of wireless products and services. The present regulation policy of spectrum utilization is to divide the spectrum into licensed lots and to allocate to different services and entities for exclusive use [29]. However, as the number of wireless devices has increased tremendously over the last few decades, the availability of wireless spectrum has become severely limited. This fact has led to a situation where new services have difficulty obtaining spectrum licenses and cannot be accommodated under the current regulation policy. This situation has been termed as "spectrum gridlock" [29] and considered as one of the main factors which thwarts further advancement and development of wireless technology.

In recent years, various strategies for overcoming spectrum gridlock have been proposed [29]. Among them, cooperative communication has been envisioned to surmount this issue. More specifically, collaborative devices can cooperate to share spectrum, time slots and resources and, ultimately, lead to more efficient communications. Upon the existing literatures, the *cognitive interference channel* (*CIFC*), depicted in Fig. 2 of [27], has been one of the most intensively studied collaborative channel model these years. Similar to the *classical interference channel* [7], [26], the CIFC is a two-user, *primary user* (*PU*) and *cognitive user* (*CU*), channel model where the two senders attempt to communicate with their respective destinations through the common medium simultaneously. The distinction between the two models lies in the fact that there is a cooperation mechanism between the two senders in the CIFC: the sender of CU has full knowledge of the message of the sender of PU. This channel model was firstly proposed and investigated by Devroye et al. [12], [16] that presented the first achievable rate region for the *discrete memoryless CIFC* (*DM-CIFC*). More significantly, their work further demonstrated that by establishing this cooperation mechanism, termed as *genie aided cognition*[1] originally, the interference caused at the destination of CU can be effectively mitigated via *Gel'fand-Pinsker precoding* (*GP precoding*) [6]. As a result, the achievable rate region for the CIFC is significantly larger than that for the classical interference channel where, contrary to the cooperative communication in the CIFC, two senders work independently. Since then, the CIFC has attracted great attentions, and several achievable rate regions for the DM-CIFC have been established [14], [17], [22]–[24], [27], [28, Section II], [34], [35], [38]. Although the capacity region for the CIFC in general is still an open problem, the capacity results for some special classes have been proved, and a clear summary can be found in [37, Section I]. Remarkably, Rini et al. proposed a new achievable rate region

---

[1] The genie aided cognition is also termed as "unidirectional cooperation" in [14], [22] while, in this paper and [42], this term is also used. However, it should be noticed that the cooperation mechanism of the former is non-causal or, more precisely, prior to transmission, but that of the latter is in a causal manner.

for the DM-CIFC encompassing all other known achievable rate regions [38].

Recently, many extensions and variants of the original CIFC have also been under researches. For instance, in [30], the channel model is basically the same as the original CIFC. However, an additional decoding requirement is imposed on the decoder of CU: to decode and understand both PU and CU's messages, and, furthermore, the secrecy level of CU's message is also taken into consideration. For this special setting, the capacity region was proved for both with and without secrecy constraint. For another instance, a more realistic model of cognition where genie-aided cognition in the original CIFC is replaced with unidirectional or bidirectional link of finite capacity between the two senders has also been under researches. This class of channel models is termed as the *interference channel with conferencing encoders / transmitters*. For example, the unidirectional link case is considered in [33] while the bidirectional link case is considered in [36], [39], [41]. There are actually other interesting variants of the original CIFC to get fit for more realistic communication scenarios (e.g., the "causal cognition[2]" channel model [20], [28, Section III], [32], [40] where the sender of CU accesses to a channel output and "causally" learns the information about PU), but this is not the focus of this paper.

In this work, we consider another extension of the original CIFC, the *cognitive interference channel with unidirectional destination cooperation* (*CIFC-UDC*), depicted in Fig. 1. This channel model not only adopts the cooperation mechanism in the original CIFC but also allows the secondary destination to participate in the communication between the sending and receiving sides of PU. Note that the original CIFC can be included as a special case of the CIFC-UDC by disabling the transmission of the secondary destination. This channel model was proposed in [42][3] which derived the first achievable rate region for the discrete memoryless case. The establishment of the asymmetrical relationship between the two destinations is motivated by this practical scenario: in licensed bands, CU is usually used to model unlicensed user (secondary user). As a "spectrum borrower", there are two tasks for CU. On the one hand, CU is not allowed to cause interference to license owner, PU. On the other hand, if possible, CU should provide some benefit for PU. It can be expected that if the second task is accomplished, both CU and PU will obtain their advantages through this transaction which attains to a win-win solution. In the CIFC-UDC, the benefit for PU is modeled as the capacity gains from the unidirectional directional cooperation and was demonstrated in [42]. However, this work is at best a proof of concept and is not integral enough for further understanding of the CIFC-UDC. For instance, the derivation of the achievable rate region missed considering the broadcasting and relaying characteristics of the CIFC-UDC. More specifically, as suggested in [28], with full knowledge of the message of PU, the sender of CU can broadcast part of the message of PU. Unfortunately, this assistance modality was missed. Besides, as a network involves relay, in general, a quantized version of the received signal at relay may be forwarded so as to improve the overall throughput [13]. Nevertheless, this coding scheme was also missed being integrated.

From an information-theoretic perspective, the CIFC-UDC is a "compound channel model" which includes several existing models, the *partially cooperative relay broadcast channel* (*PCRBC*) [19], [21], [31] and the CIFC, etc (e.g., by choosing the output of destination 2 to be null, the CIFC-UDC is degraded to the CIFC.). Although the capacity regions for the two channels are still unknown, the largest known achievable rate region for the discrete memoryless PCRBC appears in [31, Theorem 2], and that for the discrete memoryless CIFC can be found in [38, Theorem 1]. However, there lacks for a unified theorem to connect the two results, and, as a matter of fact, our main contribution is the establishment of a new unified achievable rate region for the CIFC-UDC which subsumes these two results. We note that showing the inclusion of [31, Theorem 2] implies that *Marton region* for the *broadcast channel* (*BC*) [4] is contained (see *Remark 2* in [31]). More importantly, a potentially larger achievable rate region for the PCRBC is also derived. Furthermore, the largest known achievable rate for the single-relay channel, *CMG rate* [15], and the region for the CIFC-UDC formerly derived by Chu [42, Theorem 1] are also included in this new region.

On the other hand, it is generally known that interference occurs whenever multiple users access to the common medium simultaneously, and if not well treated, interference may undermine reliable communication much more severely than noise does. Thus, in this work, how to assist the decoder of PU to alleviate the interference caused by CU via the unidirectional destination cooperation is also investigated. In fact, this can be considered as the complementary idea of the interference mitigation through GP precoding in the CIFC proposed by Devroye et al. [12], [16]. Hence, by combing these two ideas, the interferences caused at both the primary and secondary destinations will be mitigated.

The rest of this paper is structured as follows. In Section II, the notational conventions and the channel model of the CIFC-UDC are defined. In Section III, we present our new unified achievable rate region for the CIFC-UDC. Additionally, a special communication setting is employed to state the operation of the interference mitigation via the unidirectional destination cooperation. In Section IV, an outer bound is presented, and it is proved that for a special class of the CIFC-UDCs, our achievable rate region and this outer bound coincide together, resulting in the characterization of the capacity region for this class. In Section V, we focus our attention on some degradations of our achievable rate region, compare them with other existing results and prove the inclusions of [4, Theorem 2], [15], [31, Theorem 2], [38, Theorem 1] and [42, Theorem 1] and a potentially larger achievable rate region for the PCRBC. In Section VI, we conclude this paper. Lastly, Appendix presents the detailed proofs of some lemmas and theorems.

## II. NOTATIONS, DEFINITIONS AND CHANNEL MODEL

### A. Notations

The following notational conventions will be adopted throughout this paper. We use capital letters to denote the random variables and lower case letters to denote their corres-ponding realizations. We adopt the notational convenience $p_{Y|X}(y|x) = p(y|x)$ to drop the subscript of probability distribution. The expectation operation is represented as $E[\cdot]$. We write $X \sim p(x)$ to indicate that random variable $X$ is drawn according to $p(x)$, and we use $x^n$ to represent the vector $(x_1, x_2, ..., x_n)$. Moreover, $X - Y - Z$ denotes that $X$, $Y$ and $Z$ form a Markov chain. The notation $|(\cdot)|$ is used to denote the cardinality of a set. For the information theoretic quantities such as entropy, mutual information, etc, we follow the notations defined in [3].

---

[2]This setting is also termed as the *interference channel with generalized feedback* (e.g., [20], [40]) in the literatures.

[3]In [**42**], the CIFC-UDC was called as the *interference channel with degraded message sets with unidirectional destination cooperation* (*IC-DMS-UDC*).

*B. Definitions and Channel Model*

The cognitive interference channel with unidirectional destination cooperation, depicted in Fig. 1, is a channel model where sender 1 sends a message $w_1 \in W_1 = \{1,2,...,|W_1|\}$ to its destination with the help of destination 2 in n channel transmissions, and sender 2, with full knowledge of the message of sender 1, sends a message $w_2 \in W_2 = \{1,2,...,|W_2|\}$ to its destination in n channel transmissions. Additionally, we focus on the *discrete memoryless CIFC-UDC* (*DM-CIFC-UDC*). The CIFC-UDC is said to be discrete memoryless in the sense that

$$p(y_1^n, y_2^n | x_1^n, x_2^n, x_3^n) = \prod_{t=1}^{n} p(y_{1_t}, y_{2_t} | x_{1_t}, x_{2_t}, x_{3_t}) \quad \text{(II.1)}$$

for every discrete time instant t synchronously.

*Definition 1:* The DM-CIFC-UDC is described by a tuple $(\mathcal{X}_1, \mathcal{X}_2, \mathcal{X}_3, \mathcal{Y}_1, \mathcal{Y}_2, p(y_1, y_2|x_1, x_2, x_3))$, where $\mathcal{X}_1$, $\mathcal{X}_2$ and $\mathcal{X}_3$ denote the channel input alphabets, $\mathcal{Y}_1$ and $\mathcal{Y}_2$ denote the channel output alphabets, and $p(y_1, y_2|x_1, x_2, x_3)$ denotes the transition probability. Moreover, $x_1$, $x_2$ and $x_3$ are channel inputs from sender 1, sender 2 and destination 2 respectively. $y_1$ and $y_2$ denote the channel outputs at destination 1 and destination 2 respectively.

Next we present the following definitions with regard to the existence of code, the achievable rate region and the capacity region for the DM-CIFC-UDC.

*Definition 2:* A $(|W_1|, |W_2|, n, P_e)$ code consists of:

- The message set $W_k = \{1,2,...,|W_k|\}$ where $|W_k| = 2^{nR_k}$, $k = 1,2$. It is assumed that the messages $W_1$ and $W_2$ are independent and uniformly distributed.
- An encoding function for sender 1 to map $w_1 \in W_1$ to a codeword $x_1^n$:

$$f_1: w_1 \to x_1^n \quad \text{(II.2)}$$

- An encoding function for sender 2 to map $w_1 \times w_2 \in W_1 \times W_2$ to a codeword $x_2^n$:

$$f_2: w_1 \times w_2 \to x_2^n \quad \text{(II.3)}$$

- A family of encoding functions for destination 2 to map the preceding observations to the next transmitted symbol $x_{3_i}$:

$$f_{3_i}: (y_{2_1}, y_{2_2}, y_{2_3}, ..., y_{2_{i-1}}) \to x_{3_i} \text{ for } 1 \leq i \leq n \quad \text{(II.4)}$$

- The decoding function for decoder k to map $y_k^n$ to $\hat{w}_k \in W_k$:

$$g_k: y_k^n \to \hat{w}_k, k = 1,2 \quad \text{(II.5)}$$

- The average probability of error:

$$P_e \equiv \max\{P_{e_1}, P_{e_2}\} \quad \text{(II.6)}$$

where $P_{e_k}$ denotes the average probability of error of decoder k, $k = 1,2$. Furthermore, because it is assumed that the message pair $(w_1, w_2) \in (W_1, W_2)$ is equiprobable, $P_{e_k}$ can be computed as

$$P_{e_k} = \frac{1}{2^{n(R_1+R_2)}} \sum_{(w_1,w_2)} \Pr\{\hat{w}_k \neq w_k | (w_1, w_2) \text{ sent.}\} \quad \text{(II.7)}$$

*Definition 3:* A nonnegative rate pair $(R_1, R_2)$ is said to be achievable for the CIFC-UDC if there exists a $(|W_1|, |W_2|, n, P_e)$ code with $R_1 \leq \frac{1}{n} \log |W_1|$ and $R_2 \leq \frac{1}{n} \log |W_2|$ such that $P_e \to 0$ as $n \to \infty$.

*Definition 4:* The capacity region for the CIFC-UDC, denoted as $\mathcal{C}$, is the closure of the region of all the achievable rate pairs $(R_1, R_2)$. An achievable rate region, denoted as $\mathcal{R}$, is a subset of the capacity region.

### III. A New Unified Achievable Rate Region for the Discrete Memoryless Cognitive Interference Channel with Unidirectional Destination Cooperation

Since the DM-CIFC-UDC is a compound channel model with a variety of aspects, it can be expected that applying the coding schemes used in other literatures will be helpful to derive our result. The following schemes are employed in the derivation of our achievable rate region.

- **Rate-Splitting:**

*Rate-splitting* was developed in [7] to derive *HK region* for the classical interference channel and, since then, has been widely employed to derive capacity results for multi-user networks. In our work, the message of user 2 is split into public and private sub-messages, $W_{2P}$ and $W_{22}$, and the message of user 1 is split into public, private and broadcast sub-messages, $W_{1P}$, $W_{11}$ and $W_{1B}$ respectively. The purpose for splitting a message into public and private sub-messages is to enable the unintended destination to jointly decode part of the message sent from another user and, equivalently, to eliminate the interference associated with this part of the message. On the other hand, the broadcast sub-message, $W_{1B}$, allows sender 2 to broadcast part of the message of user 1. This technique was formerly used to expand the achievable rate region for the CIFC further (e.g., [28], [38]) and also employed to derive an achievable rate region for *the broadcast channel with two cognitive relays* [35].

- **Gel'fand-Pinsker Binning / Precoding:**

GP precoding and, in particular, *writing on dirty paper coding* (*WDPC*) [10] in the Gaussian channel enable encoder to precode so as to mitigate the "non-causally" known interference. Although the term non-causally seems to be physically infeasible in one's common sense, in the CIFC and CIFC-UDC also, the message of PU is assumed to be known to the sender of CU ahead of time (The viability and reasonability of this assumption are discussed in [29, Section V].). As a result, this prior knowledge allows the sender of CU to perform GP precoding to mitigate the interference caused at its destination.

Remarkably, in the single-user Gaussian channel, WDPC enables the encoder to precode its message at the rate associated with "interference-free communication" [10].

- **Marton Binning:**

For the broadcast channel, Marton region [4, Theorem 2], derived by *Marton binning*, is the largest known achievable rate region. Due to the broadcasting nature of the CIFC-UDC, in our coding scheme, Marton binning is employed at sender 2 to broadcast $W_{1B}$ and $W_{22}$ to destination 1 and destination 2 respectively.

- **Cooperative Strategy I: Generalized Strategy of Cover and El Gamal:**

The two fundamental cooperative coding schemes for the single-relay channel are *decode-and-forward* (*DF*) [5, Theorem 1] and *compress-and-forward* (*CF*) [5, Theorem 6] strategies. For DF strategy, the relay decodes and understands all the information sent from the sender and forwards this information to the destination while, for CF strategy, the relay decodes none of the sender's message but forwards a quantized version of its received signals to the destination. Remarkably, for the Gaussian single-relay channel, DF strategy achieves the capacity when the relay is at the sender while CF strategy achieves the capacity when the relay is at the destination [3, Section 9.2.1]. Nevertheless, to date, the capacity remains unsolved between the two extreme cases.

These two schemes were further combined and generalized in [5, Theorem 7] which is known as *generalized strategy of Cover and El Gamal*. In our coding scheme of the unidirectional destination cooperation, the similar strategy as generalized strategy of Cover and El Gamal is taken to derive our result.

- **Cooperative Strategy II: Interference Forwarding Technique:**

*Interference forwarding technique* (*IFT*) indicates that, in relay networks, not only does forwarding wanted message to intended destination help the overall throughput but also forwarding unwanted message to unintended destination does. This technique was proposed by Mari'c et al. [25], and the authors considered an interesting scenario (see Fig. 2 in [25]) to further demonstrate this idea. In the following, we employ this scenario and restate how interference forwarding technique functions.

The *interference channel with a relay* (*ICR*), depicted in Fig. 1 of [25], is considered, and an additional assumption is imposed.

*Assumption*:

$$p(y_3|x_1, x_2, x_3) = p(y_3|x_2, x_3) \quad (\text{III. 1})$$

Equivalently, the relay cannot observe $x_1$. Henceforth, the relay cannot forward any information about $W_1$. In spite of this assumption, the relay still can help both the destinations and improve the throughput of this two-user network. To be more specific, though relaying the unwanted message $W_2$ to destination 1 is only "forwarding interference" from the aspect of destination 1, that indeed enhances the reception of $W_2$ at destination 1. As a result, the relay can effectively place destination 1 in the "strong interference regime" which allows destination 1 to decode $W_2$ and to eliminate the interference associated with this message. In Fig. 5 of [25], a numerical analysis is provided to demonstrate the improvement.

In our work, we integrate IFT into the secondary destination. That is, the secondary destination not only forwards the wanted message $W_{1P}$ to destination 1 but also forwards $W_{2P}$ on to alleviate the interference caused by CU's transmission at the destination of PU. This cooperation modality serves as an important role in a class of Gaussian CIFC-UDCs presented below. More interestingly, by combing IFT with GP precoding, the interferences caused at both the destinations can be alleviated, which is impossible to be attained by the strategy of [42, Theorem 1].

- **The Improved SimBack Decoding:**

The *backward decoding* was introduced by Willems for the work on the *multiple access channel with generalized feedback* (*MAC-GF*) [9, Chapter 7]. Later, Chong et al. proposed the *improved simback decoding* and made use of generalized relaying strategy of Cover and El Gamal to prove the new achievable rate, *CMG rate*, for the single-relay channel [15]. In this work, a similar decoding strategy as the improved simback decoding is employed by decoder 1 to derive our result.

In the following, we present our new unified achievable rate region for the DM-CIFC-UDC, the main result of this paper.

*Theorem 1:* Let $\mathcal{P}$ denote the set of all joint probability distributions

$$p(u_{1p}, u_1, v_1, u_{2p}, u_2, v_{12}, v_2, x_1, x_2, x_3, y_1, y_2, \hat{y}_2) \quad (\text{III. 2. A})$$

which can be factored in the following form:

$$\begin{aligned}
&p(u_{1p})p(u_1|u_{1p})p(v_1|u_{1p}, u_1)p(u_{2p}|u_{1p}) \\
&\times p(u_2, v_{12}, v_2|u_{1p}, u_1, v_1, u_{2p})p(x_1|u_{1p}, u_1, v_1) \\
&\times p(x_2|u_{1p}, u_1, v_1, u_{2p}, u_2, v_{12}, v_2)p(x_3|u_{1p}, u_{2p}) \\
&\times p(y_1, y_2|x_1, x_2, x_3)p(\hat{y}_2|u_{1p}, u_1, u_{2p}, u_2, x_3, y_2)
\end{aligned} \quad (\text{III. 2. B})$$

Let $\mathcal{R}(p)$ be the set of all nonnegative rate pairs $(R_1, R_2) = (R_{11} + R_{1P} + R_{1B}, R_{22} + R_{2P})$ such that the following constraints hold:

$$\begin{aligned}
R_{11} \geq 0, R_{1P} \geq 0, R_{1B} \geq 0, \\
R_{22} \geq 0, R_{2P} \geq 0
\end{aligned} \quad (\text{III. 2. C})$$

$$\begin{aligned}
R'_{2P} &\geq A \\
R'_{1B} &\geq 0 \\
R'_{22} &\geq I(V_1; V_2|U_{1p}, U_1, U_{2p}, U_2) \\
R'_{1B} + R'_{22} &\geq I(V_1, V_{12}; V_2|U_{1p}, U_1, U_{2p}, U_2)
\end{aligned} \quad (\text{III. 2. D})$$

$$\begin{aligned}
R_{1p} + R_{11} + L_{2P} + L_{1B} &\leq A + B + D - C \\
R_{11} + L_{2P} + L_{1B} &\leq A + B + E - C \\
L_{2P} + L_{1B} &\leq A + B + H - C \\
R_{11} + L_{1B} &\leq \min\binom{A + B + F - C,}{A + G} \\
L_{1B} &\leq \min(J, B + I - C)
\end{aligned} \quad (\text{III. 2. E})$$

$$\begin{aligned}
R_{1P} + L_{2P} + L_{22} &\leq K \\
L_{2P} + L_{22} &\leq L \\
L_{22} &\leq M
\end{aligned} \quad (\text{III. 2. F})$$

$$C \leq I(Y_1; X_3|U_{1p}, U_1, V_1, U_{2p}, U_2, V_{12}) + B \quad (\text{III. 2. G})$$

TABLE I
ENCODING AND DECODING MESSAGE INDICES ALONG THE BLOCKS FOR
THE COGNITIVE INTERFERENCE CHANNEL WITH UNIDIRECTIONAL DESTINATION COOPERATION

| block 1 | ... | block b | ... | block (B + 1) | block (B + 2) |
|---|---|---|---|---|---|
| $u_{1p}^n(1)$ | ... | $u_{1p}^n(j_p^{(b)})$ | ... | $u_{1p}^n(j_p^{(B+1)})$ | $u_{1p}^n(1)$ |
| $u_1^n(j^{(1)}\|1)$ | ... | $u_1^n(j^{(b)}\|j_p^{(b)})$ | ... | $u_1^n(1\|j_p^{(B+1)})$ | $u_1^n(1\|1)$ |
| $v_1^n(1\|1, j^{(1)})$ | ... | $v_1^n(i_p^{(b)}\|j_p^{(b)}, j^{(b)})$ | ... | $v_1^n(i_p^{(B+1)}\|j_p^{(B+1)}, 1)$ | $v_1^n(1\|1,1)$ |
| $u_{2p}^n(1\|1)$ | ... | $u_{2p}^n(m_p^{(b)}\|j_p^{(b)})$ | ... | $u_{2p}^n(m_p^{(B+1)}\|j_p^{(B+1)})$ | $u_{2p}^n(1\|1)$ |
| $u_2^n(m^{(1)}, m^{*(1)}\|1, j^{(1)}, 1)$ | ... | $u_2^n(m^{(b)}, m^{*(b)}\|j_p^{(b)}, j^{(b)}, m_p^{(b)})$ | ... | $u_2^n(1, m^{*(B+1)}\|j_p^{(B+1)}, 1, m_p^{(B+1)})$ | $u_2^n(1, m^{*(B+2)}\|1,1,1)$ |
| $v_{12}^n\begin{pmatrix}1, & 1, j^{(1)}, 1, \\ k^{*(1)} & 1, m^{(1)}, m^{*(1)}\end{pmatrix}$ | ... | $v_{12}^n\begin{pmatrix}k_p^{(b)}, & j_p^{(b)}, j^{(b)}, i_p^{(b)}, \\ k^{*(b)} & m_p^{(b)}, m^{(b)}, m^{*(b)}\end{pmatrix}$ | ... | $v_{12}^n\begin{pmatrix}k_p^{(B+1)}, & j_p^{(B+1)}, 1, i_p^{(B+1)}, \\ k^{*(B+1)} & m_p^{(B+1)}, 1, m^{*(B+1)}\end{pmatrix}$ | $v_{12}^n\begin{pmatrix}1, & 1,1,1, \\ k^{*(B+2)} & 1, 1, m^{*(B+2)}\end{pmatrix}$ |
| $v_2^n\begin{pmatrix}l^{(1)}, & 1, j^{(1)}, 1, \\ l^{*(1)} & m^{(1)}, m^{*(1)}\end{pmatrix}$ | ... | $v_2^n\begin{pmatrix}l^{(b)}, & j_p^{(b)}, j^{(b)}, m_p^{(b)}, \\ l^{*(b)} & m^{(b)}, m^{*(b)}\end{pmatrix}$ | ... | $v_2^n\begin{pmatrix}1, & j_p^{(B+1)}, 1, m_p^{(B+1)}, \\ l^{*(B+1)} & 1, m^{*(B+1)}\end{pmatrix}$ | $v_2^n\begin{pmatrix}1, & 1,1,1, \\ l^{*(B+2)} & 1, 1, m^{*(B+2)}\end{pmatrix}$ |
| $x_1^n(1, j^{(1)}, 1)$ | ... | $x_1^n(j_p^{(b)}, j^{(b)}, i_p^{(b)})$ | ... | $x_1^n(j_p^{(B+1)}, 1, i_p^{(B+1)})$ | $x_1^n(1,1,1)$ |
| $x_2^n\begin{pmatrix}1, j^{(1)}, 1, \\ 1, m^{(1)}, 1, l^{(1)}\end{pmatrix}$ | ... | $x_2^n\begin{pmatrix}j_p^{(b)}, j^{(b)}, i_p^{(b)}, \\ m_p^{(b)}, m^{(b)}, k_p^{(b)}, l^{(b)}\end{pmatrix}$ | ... | $x_2^n\begin{pmatrix}j_p^{(B+1)}, 1, i_p^{(B+1)}, \\ m_p^{(B+1)}, 1, k_p^{(B+1)}, 1\end{pmatrix}$ | $x_2^n\begin{pmatrix}1,1,1, \\ 1,1,1,1\end{pmatrix}$ |
| $x_3^n(1\|1,1)$ | ... | $x_3^n(z_p^{(b)}\|j_p^{(b)}, m_p^{(b)})$ | ... | $x_3^n(z_p^{(B+1)}\|j_p^{(B+1)}, m_p^{(B+1)})$ | $x_3^n(z_p^{(B+2)}\|1,1)$ |
| [— — —] | ... | $(\hat{j}_{PD_1}^{(b)}, \hat{i}_{PD_1}^{(b)}, \hat{m}_{PD_1}^{(b)}, \hat{k}_{PD_1}^{(b)}, \hat{z}_{PD_1}^{(b)})$ | ... | $\begin{pmatrix}\hat{j}_{PD_1}^{(B+1)}, \hat{i}_{PD_1}^{(B+1)}, \hat{m}_{PD_1}^{(B+1)}, \\ \hat{k}_{PD_1}^{(B+1)}, \hat{z}_{PD_1}^{(B+1)}\end{pmatrix}$ | [Initiation][4] |
| $(\hat{j}_{D2}^{(1)}, \hat{m}_{D2}^{(1)}, \hat{l}_{D2}^{(1)}, \hat{z}_{D2}^{(1)})$ | ... | $(\hat{j}_{D2}^{(b)}, \hat{m}_{D2}^{(b)}, \hat{l}_{D2}^{(b)}, \hat{z}_{D2}^{(b)})$ | ... | $\begin{pmatrix}\hat{j}_{D2}^{(B+1)} = 1, \hat{m}_{D2}^{(B+1)} = 1, \\ \hat{l}_{D2}^{(B+1)} = 1, \hat{z}_{D2}^{(B+1)}\end{pmatrix}$ | [— — —] |

\* The second row shows the sequences constructed by the auxiliary random variables and the sub-messages conveyed by them.
\* The third row shows the transmitted codewords in the corresponding block.
\* The fourth row shows the decoding results evaluated after the corresponding block by decoder 1 and decoder 2 respectively.
\* Some indices are initially set to be equal to 1 due to mathematical symmetry.

[4] Refer to the first decoding step in **[The Decoding Process of Decoder 1]**.

where

$$L_{1B} = R_{1B} + R'_{1B}, \quad L_{2P} = R_{2P} + R'_{2P}, \quad L_{22} = R_{22} + R'_{22}$$

and

$$\begin{aligned}
A &= I(V_1; U_2|U_{1p}, U_1, U_{2p}) \\
B &= I(Y_1, V_1, V_{12}; \hat{Y}_2|U_{1p}, U_1, U_{2p}, U_2, X_3) \\
C &= I(Y_2; \hat{Y}_2|U_{1p}, U_1, U_{2p}, U_2, X_3) \\
D &= I(Y_1; U_{1p}, U_1, V_1, U_{2p}, U_2, V_{12}, X_3) \\
E &= I(Y_1; V_1, U_{2p}, U_2, V_{12}, X_3|U_{1p}, U_1) \\
F &= I(Y_1; V_1, V_{12}, X_3|U_{1p}, U_1, U_{2p}, U_2) \\
G &= I(Y_1, \hat{Y}_2; V_1, V_{12}|U_{1p}, U_1, U_{2p}, U_2, X_3) \\
H &= I(Y_1; U_{2p}, U_2, V_{12}, X_3|U_{1p}, U_1, V_1) \\
I &= I(Y_1; V_{12}, X_3|U_{1p}, U_1, V_1, U_{2p}, U_2) \\
J &= I(Y_1, \hat{Y}_2; V_{12}|U_{1p}, U_1, V_1, U_{2p}, U_2, X_3) \\
K &= I(Y_2; U_1, U_2, V_2|U_{1p}, U_{2p}, X_3) \\
L &= I(Y_2; U_2, V_2|U_{1p}, U_1, U_{2p}, X_3) \\
M &= I(Y_2; V_2|U_{1p}, U_1, U_{2p}, U_2, X_3)
\end{aligned}$$

Since destination 1 (2) does not interest in $W_{2P}$ ($W_{1P}$), some rate constraints can be dropped:

- The first constraint in (III.2.E) can be dropped if $R_{1p} = R_{11} = L_{1B} = 0$.
- The second constraint in (III.2.E) can be dropped if $R_{11} = L_{1B} = 0$.
- The third constraint in (III.2.E) can be dropped if $L_{1B} = 0$.
- The first constraint in (III.2.F) can be dropped if $L_{2P} = L_{22} = 0$.

Then the region $\mathcal{R} = \cup_{p(.) \in \mathcal{P}} \mathcal{R}(p)$ is an achievable rate region for the DM-CIFC-UDC.

*Proof:* For separately encoding, the messages $W_1$ of $nR_1$ bits and $W_2$ of $nR_2$ bits are split as follows:

$$\begin{aligned}
W_1 &= (W_{11}, W_{1P}, W_{1B}) & \text{(III.3.A)} \\
W_2 &= (W_{22}, W_{2P}) & \text{(III.3.B)}
\end{aligned}$$

where the sub-message $W_i$ is $nR_i$ bits, $i \in \{11, 1P, 1B, 22, 2P\}$.

Hence, the transmission rate pair $(R_1, R_2)$ is

$$\begin{aligned}
R_1 &= R_{11} + R_{1P} + R_{1B} & \text{(III.4.A)} \\
R_2 &= R_{22} + R_{2P} & \text{(III.4.B)}
\end{aligned}$$

The purpose of each sub-message is illustrated as follows:

- $W_{11}$ and $W_{22}$ are the private sub-messages of PU and CU which will be decoded by the intended decoder only and treated as noise at the unintended destination.
- $W_{1P}$ and $W_{2P}$ are the public sub-messages of PU and CU which will be decoded by both the decoders.
- $W_{1B}$ is the broadcast sub-message of PU. This sub-message will be broadcasted by the sender of CU and decoded by the decoder of PU only.

Next, let us consider the coding scheme employed in this work. Our coding scheme is a block Markov scheme, and its flow is shown schematically in Table I. We consider $B + 2$ blocks, each of $n$ symbols. A sequence of $B$ sub-message tuples $(i^{(b)}, j^{(b)}, k^{(b)}, l^{(b)}, m^{(b)})$, $i^{(b)} \in \{1, ..., 2^{nR_{11}}\}$, $j^{(b)} \in \{1, ..., 2^{nR_{1P}}\}$, $k^{(b)} \in \{1, ..., 2^{nR_{1B}}\}$, $l^{(b)} \in \{1, ..., 2^{nR_{22}}\}$ and $m^{(b)} \in \{1, ..., 2^{nR_{2P}}\}$, $b = 1, ..., B$, will be sent over the channel in $n \times (B + 2)$ symbols (or, equivalently, $B + 2$ blocks). Note that as $B \to \infty$, for fixed n, the rate pair $(R_1, R_2) = (\frac{B(R_{11}+R_{1P}+R_{1B})}{B+2}, \frac{B(R_{22}+R_{2P})}{B+2})$ is arbitrarily close to $(R_{11} + R_{1P} + R_{1B}, R_{22} + R_{2P})$.

In the following, codebook construction, encoding and transmission, decoding and probability of error analysis will be presented sequentially. At first, we note the notations followed in this proof. The index $i_p^{(b)}$ represents the index of message in the previous block—i.e., $i_p^{(b)} = i^{(b-1)}$. In addition, the subscript of the index $\hat{i}_{D1}^{(b)}$ ($\hat{i}_{D2}^{(b)}$) denotes the decoding result evaluated by decoder 1 (2).

**Codebook Construction:**

In each block b, $b = 1, \ldots, B + 2$, we shall use the codebook constructed as below:

- Generate $2^{nR_{1P}}$ n-sequences $u_{1p}^n$, each with probability

$$p(u_{1p}^n) = \prod_{t=1}^n p(u_{1p_t}).$$

Label them as $u_{1p}^n(j_p)$, where $j_p \in \{1, \ldots, 2^{nR_{1P}}\}$.

- For each $u_{1p}^n(j_p)$, generate $2^{nR_{1P}}$ n-sequences $u_1^n$, each with probability

$$p(u_1^n|u_{1p}^n(j_p)) = \prod_{t=1}^n p(u_{1_t}|u_{1p_t}(j_p)).$$

Label them as $u_1^n(j|j_p)$, where $j \in \{1, \ldots, 2^{nR_{1P}}\}$.

- For each $(u_{1p}^n(j_p), u_1^n(j|j_p))$, generate $2^{nR_{11}}$ n-sequences $v_1^n$, each with probability

$$p(v_1^n|u_{1p}^n(j_p), u_1^n(j|j_p)) = \prod_{t=1}^n p(v_{1_t}|u_{1p_t}(j_p), u_{1_t}(j|j_p)).$$

Label them as $v_1^n(i_p|j_p, j)$, where $i_p \in \{1, \ldots, 2^{nR_{11}}\}$.

- For each $u_{1p}^n(j_p)$, generate $2^{nR_{2P}}$ n-sequences $u_{2p}^n$, each with probability

$$p(u_{2p}^n|u_{1p}^n(j_p)) = \prod_{t=1}^n p(u_{2p_t}|u_{1p_t}(j_p)).$$

Label them as $u_{2p}^n(m_p|j_p)$, where $m_p \in \{1, \ldots, 2^{nR_{2P}}\}$.

- For each $(u_{1p}^n(j_p), u_1^n(j|j_p), u_{2p}^n(m_p|j_p))$, generate $2^{n(R_{2P}+R'_{2P})}$ n-sequences $u_2^n$, each with probability

$$p(u_2^n|u_{1p}^n(j_p), u_1^n(j|j_p), u_{2p}^n(m_p|j_p))$$
$$= \prod_{t=1}^n p(u_{2_t}|u_{1p_t}(j_p), u_{1_t}(j|j_p), u_{2p_t}(m_p|j_p)).$$

Label them as $u_2^n(m, m'|j_p, j, m_p)$, where $m \in \{1, \ldots, 2^{nR_{2P}}\}$ and $m' \in \{1, \ldots, 2^{nR'_{2P}}\}$.

- For each $\begin{pmatrix} u_{1p}^n(j_p), u_1^n(j|j_p), v_1^n(i_p|j_p, j), \\ u_{2p}^n(m_p|j_p), u_2^n(m, m'|j_p, j, m_p) \end{pmatrix}$, generate $2^{n(R_{1B}+R'_{1B})}$ n-sequences $v_{12}^n$, each with probability

$$p(v_{12}^n|u_{1p}^n(j_p), u_1^n(j|j_p), v_1^n(i_p|j_p, j), u_{2p}^n(m_p|j_p), u_2^n(m, m'|j_p, j, m_p))$$
$$= \prod_{t=1}^n p\left(v_{12_t} \middle| \begin{matrix} u_{1p_t}(j_p), u_{1_t}(j|j_p), v_{1_t}(i_p|j_p, j), \\ u_{2p_t}(m_p|j_p), u_{2_t}(m, m'|j_p, j, m_p) \end{matrix}\right).$$

Label them as $v_{12}^n(k_p, k'|j_p, j, i_p, m_p, m, m')$, where $k_p \in \{1, \ldots, 2^{nR_{1B}}\}$ and $k' \in \{1, \ldots, 2^{nR'_{1B}}\}$.

- For each $\begin{pmatrix} u_{1p}^n(j_p), u_1^n(j|j_p), u_{2p}^n(m_p|j_p), \\ u_2^n(m, m'|j_p, j, m_p) \end{pmatrix}$, generate $2^{n(R_{22}+R'_{22})}$ n-sequences $v_2^n$, each with probability

$$p(v_2^n|u_{1p}^n(j_p), u_1^n(j|j_p), u_{2p}^n(m_p|j_p), u_2^n(m, m'|j_p, j, m_p))$$
$$= \prod_{t=1}^n p(v_{2_t}|u_{1p_t}(j_p), u_{1_t}(j|j_p), u_{2p_t}(m_p|j_p), u_{2_t}(m, m'|j_p, j, m_p)).$$

Label them as $v_2^n(l, l'|j_p, j, m_p, m, m')$, where $l \in \{1, \ldots, 2^{nR_{22}}\}$ and $l' \in \{1, \ldots, 2^{nR'_{22}}\}$.

Here we apply Gel'fand-Pinsker binning and Marton binning to further construct our codebook.

- **[Gel'fand-Pinsker Binning]**

For fixed $(j_p, j, i_p, m_p, m)$, let encoder 2 find an $m'$ such that

$$\left\{ \begin{pmatrix} u_{1p}^n(j_p), u_1^n(j|j_p), v_1^n(i_p|j_p, j), \\ u_{2p}^n(m_p|j_p), u_2^n(m, m'|j_p, j, m_p) \end{pmatrix} \in T_\epsilon^n \right\} \quad \text{(III. 5)}$$

It is proved in lemma 1, provided in Appendix (A), that with sufficiently high probability, encoder 2 can find at least one such $m'$ provided that

$$R'_{2P} \geq A \quad \text{(III. 6)}$$

and n is sufficiently large. Certainly, this $m'$ can be assigned as a function of $(j_p, j, i_p, m_p, m)$. We denote this $m'$ as $m^* = f^*_{GP}(j_p, j, i_p, m_p, m)$.

- **[Marton Binning]**

For fixed $(j_p, j, i_p, m_p, m, k_p, l)$ and previously found $m^*$, let encoder 2 further find a pair $(k', l')$ such that

$$\left\{ \begin{pmatrix} u_{1p}^n(j_p), u_1^n(j|j_p), v_1^n(i_p|j_p, j), \\ u_{2p}^n(m_p|j_p), u_2^n(m, m^*|j_p, j, m_p), \\ v_{12}^n(k_p, k'|j_p, j, i_p, m_p, m, m^*), \\ v_2^n(l, l'|j_p, j, m_p, m, m^*) \end{pmatrix} \in T_\epsilon^n \right\} \quad \text{(III. 7)}$$

It is proved in lemma 2, presented in Appendix (B), that with sufficiently high probability, encoder 2 can find at least one such pair $(k', l')$ provided that

$$R'_{1B} \geq 0 \quad \text{(III. 8. A)}$$
$$R'_{22} \geq I(V_1; V_2|U_{1p}, U_1, U_{2p}, U_2) \quad \text{(III. 8. B)}$$
$$R'_{1B} + R'_{22} \geq I(V_1, V_{12}; V_2|U_{1p}, U_1, U_{2p}, U_2) \quad \text{(III. 8. C)}$$

and n is sufficiently large. Certainly, this pair $(k', l')$ can be assigned as a function of $(j_p, j, i_p, m_p, m, k_p, l)$. We denote this pair $(k', l')$ as $(k^*, l^*) = f^*_M(j_p, j, i_p, m_p, m, k_p, l)$.

Now let us generate the codebooks for sender 1, sender 2 and destination 2.

- **[Sender 1]**

For each $(u_{1p}^n(j_p), u_1^n(j|j_p), v_1^n(i_p|j_p, j))$, generate an n-sequence $x_1^n$ with probability

$$p(x_1^n|u_{1p}^n(j_p), u_1^n(j|j_p), v_1^n(i_p|j_p, j))$$
$$= \prod_{t=1}^n p(x_{1_t}|u_{1p_t}(j_p), u_{1_t}(j|j_p), v_{1_t}(i_p|j_p, j)).$$

Label it as $x_1^n(j_p, j, i_p)$.

- **[Sender 2]**

  For each $\begin{pmatrix} u_{1p}^n(j_p), u_1^n(j|j_p), v_1^n(i_p|j_p, j), \\ u_{2p}^n(m_p|j_p), u_2^n(m, m^*|j_p, j, m_p), \\ v_{12}^n(k_p, k^*|j_p, j, i_p, m_p, m, m^*), \\ v_2^n(l, l^*|j_p, j, m_p, m, m^*) \end{pmatrix}$, generate an n-sequence $x_2^n$ with probability

  $$p\left(x_2^n \middle| \begin{array}{l} u_{1p}^n(j_p), u_1^n(j|j_p), v_1^n(i_p|j_p, j), \\ u_{2p}^n(m_p|j_p), u_2^n(m, m^*|j_p, j, m_p), \\ v_{12}^n(k_p, k^*|j_p, j, i_p, m_p, m, m^*), \\ v_2^n(l, l^*|j_p, j, m_p, m, m^*) \end{array}\right)$$
  $$= \prod_{t=1}^n p\left(x_{2_t} \middle| \begin{array}{l} u_{1p_t}(j_p), u_{1_t}(j|j_p), v_{1_t}(i_p|j_p, j), \\ u_{2p_t}(m_p|j_p), u_{2_t}(m, m^*|j_p, j, m_p), \\ v_{12_t}(k_p, k^*|j_p, j, i_p, m_p, m, m^*), \\ v_{2_t}(l, l^*|j_p, j, m_p, m, m^*) \end{array}\right).$$

  Label it as $x_2^n(j_p, j, i_p, m_p, m, k_p, l)$.

- **[Destination 2]**

  For each $\left(u_{1p}^n(j_p), u_{2p}^n(m_p|j_p)\right)$, generate $2^{n\hat{R}}$ n-sequences $x_3^n$, each with probability

  $$p\left(x_3^n \middle| u_{1p}^n(j_p), u_{2p}^n(m_p|j_p)\right)$$
  $$= \prod_{t=1}^n p\left(x_{3_t} \middle| u_{1p_t}(j_p), u_{2p_t}(m_p|j_p)\right).$$

  Label them as $x_3^n(z_p|j_p, m_p)$, where $z_p \in \{1, \ldots, 2^{n\hat{R}}\}$.

  Besides, for each $\begin{pmatrix} u_{1p}^n(j_p), u_1^n(j|j_p), u_{2p}^n(m_p|j_p), \\ u_2^n(m, m^*|j_p, j, m_p), x_3^n(z_p|j_p, m_p) \end{pmatrix}$, generate $2^{n\tilde{R}}$ n-sequences $\hat{y}_2^n$, each with probability

  $$p\left(\hat{y}_2^n \middle| \begin{array}{l} u_{1p}^n(j_p), u_1^n(j|j_p), u_{2p}^n(m_p|j_p), \\ u_2^n(m, m^*|j_p, j, m_p), x_3^n(z_p|j_p, m_p) \end{array}\right)$$
  $$= \prod_{t=1}^n p\left(\hat{y}_{2_t} \middle| \begin{array}{l} u_{1p_t}(j_p), u_{1_t}(j|j_p), u_{2p_t}(m_p|j_p), \\ u_{2_t}(m, m^*|j_p, j, m_p), x_{3_t}(z_p|j_p, m_p) \end{array}\right).$$

  Label them as $\hat{y}_2^n(z|j_p, j, m_p, m, m^*, z_p)$, where $z \in \{1, \ldots, 2^{n\tilde{R}}\}$. Note that $\hat{y}_2^n$ represents a quantized version of $y_2^n$.

Our codebooks are completely constructed. Now let us move our attention to the encoding and transmission processes of sender 1, sender 2 and destination 2.

**Encoding and Transmission**

The encoding and transmission processes are performed in $(B+2)$ blocks. Note that destination 2 serves a dual purpose: to work as the receiving dual for sender 2 and to participate in the communication between the sending and receiving sides of PU. Due to this operation, the encoding and decoding processes of destination 2 are combined together. It is assumed that prior to the decoding process of block b (Notice that decoder 1 performs the backward decoding and starts its decoding process only after the entire transmission.):

- Decoder 1 has available

  $\left(j_p^{(b+1)}, j_p^{(b+2)}, \ldots, j_p^{(B+1)}\right)$,
  $\left(i_p^{(b+1)}, i_p^{(b+2)}, \ldots, i_p^{(B+1)}\right)$
  $\left(m_p^{(b+1)}, m_p^{(b+2)}, \ldots, m_p^{(B+1)}\right)$,
  $\left(k_p^{(b+1)}, k_p^{(b+2)}, \ldots, k_p^{(B+1)}\right)$ and
  $\left(z_p^{(b+1)}, z_p^{(b+2)}, \ldots, z_p^{(B+1)}\right)$, where $b = B, B-1, \ldots, 1$.

- Decoder 2 has available

  $\left(j^{(1)}, j^{(2)}, \ldots, j^{(b-1)}\right)$,
  $\left(m^{(1)}, m^{(2)}, \ldots, m^{(b-1)}\right)$,
  $\left(l^{(1)}, l^{(2)}, \ldots, l^{(b-1)}\right)$ and
  $\left(z^{(1)}, z^{(2)}, \ldots, z^{(b-1)}\right)$, where $b = 2, 3, \ldots, B-1, B$.

Sender 1, sender 2 and destination 2 send the following sequences of codewords in each block b, $b = 1, \ldots, B+2$:

- $b = 1$,
  Sender 1:
  $$x_1^n\left(j_p^{(1)} = 1, j^{(1)}, i_p^{(1)} = 1\right).$$
  Sender 2:
  $$x_2^n\left(j_p^{(1)} = 1, j^{(1)}, i_p^{(1)} = 1, m_p^{(1)} = 1, m^{(1)}, k_p^{(1)} = 1, l^{(1)}\right).$$
  Destination 2:
  $$x_3^n(1|1,1).$$

- $b = 2, \ldots, B$,
  Sender 1:
  $$x_1^n\left(j_p^{(b)}, j^{(b)}, i_p^{(b)}\right).$$
  Sender 2:
  $$x_2^n\left(j_p^{(b)}, j^{(b)}, i_p^{(b)}, m_p^{(b)}, m^{(b)}, k_p^{(b)}, l^{(b)}\right).$$
  Destination 2:
  $$x_3^n\left(z_p^{(b)} \middle| j_p^{(b)}, m_p^{(b)}\right).$$

- $b = B + 1$,
  Sender 1:
  $$x_1^n\left(j_p^{(B+1)}, 1, i_p^{(B+1)}\right).$$
  Sender 2:
  $$x_2^n\left(j_p^{(B+1)}, 1, i_p^{(B+1)}, m_p^{(B+1)}, 1, k_p^{(B+1)}, 1\right).$$
  Destination 2:
  $$x_3^n\left(z_p^{(B+1)} \middle| j_p^{(B+1)}, m_p^{(B+1)}\right).$$

- $b = B + 2$,
  Sender 1:
  $$x_1^n(1,1,1).$$
  Sender 2:
  $$x_2^n(1,1,1,1,1,1,1).$$
  Destination 2:
  $$x_3^n\left(z_p^{(B+2)} \middle| 1,1\right).$$

(The above is presented schematically in the third row of Table I.)

**Decoding**

**[The Decoding Process of Decoder 1]**

Decoder 1 employs the backward decoding and starts the decoding process only after receiving block $(B + 2)$. In addition, the decoding process is divided into two steps:

- The First Decoding Step:

    As can be observed in the last column of Table I, in order to perform the backward decoding from block $(B + 2)$, an initial condition for $z_p^{(B+2)}$ is necessary. In this step, we employ the approach in [18] to provide a proper initiation:

    For each $x_3^n\left(z_p^{(B+2)}\middle|1,1\right)$, where $z_p^{(B+2)} \in \{1, \dots, 2^{n\hat{R}}\}$, generate an n-sequence $y_2^n$ with probability

    $$p\left(y_2^n \middle| x_1^n(1,1,1), x_2^n(1,1,1,1,1,1), x_3^n\left(z_p^{(B+2)}\middle|1,1\right)\right)$$
    $$= \prod_{t=1}^{n} p\left(y_{2_t} \middle| x_{1_t}(1,1,1), x_{2_t}(1,1,1,1,1,1), x_{3_t}\left(z_p^{(B+2)}\middle|1,1\right)\right).$$

    Next, choose a $z^* \in \{1, \dots, 2^{n\hat{R}}\}$ such that

    $$\left\{\begin{pmatrix} u_{1p}^n(1), u_1^n(1|1), u_{2p}^n(1|1), \\ u_2^n\left(1, m^{*(B+2)}\middle|1,1,1\right), x_3^n\left(z_p^{(B+2)}\middle|1,1\right), \\ \hat{y}_2^n\left(z^*\middle|1,1,1,1, m^{*(B+2)}, z_p^{(B+2)}\right), y_2^n \end{pmatrix} \in T_\epsilon^n\right\} \quad \text{(III.9)}$$

    (Notice that $m^{*(B+2)}$ can be determined since $m^{*(B+2)} = f_{GP}^*\left(j_p^{(B+2)}, j^{(B+2)}, i_p^{(B+2)}, m_p^{(B+2)}, m^{(B+2)}\right)$, and all the indices are set to be equal to 1 in block $(B + 2)$ (see Table I).)

    Then, swap the labeling of

    $$\hat{y}_2^n\left(z^*\middle|1,1,1,1, m^{*(B+2)}, z_p^{(B+2)}\right)$$

    and

    $$\hat{y}_2^n\left(1\middle|1,1,1,1, m^{*(B+2)}, z_p^{(B+2)}\right)$$

    if one such $z^*$ exists, and do nothing if no such $z^*$ is found. Following the similar argument as the proof of lemma 1, we can prove that such a $z^*$ will exist with sufficiently high probability if

    $$\hat{R} \geq C \quad \text{(III.10)}$$

    and n is sufficiently large.

- The Second Decoding Step:

    After the first step, decoder 1 has a proper initiation for the backward decoding and starts the decoding process from block $(B + 1)$ backward to block B, and so on. It is assumed that decoder 1 knows $(j^{(b)}) = \left(j_p^{(b+1)}\right)$, $(m^{(b)}) = \left(m_p^{(b+1)}\right)$ and $(z^{(b)}) = \left(z_p^{(b+1)}\right)$ from block $(b + 1)$ and then considers block b, $b = B, (B - 1), \dots, 2$:

    Based on the received sequence $y_1^{n^{(b)}}$ where the superscript "(b)" notes that the observation is obtained from block b, decoder 1 chooses a tuple $\left(\hat{j}_{p_{D_1}}^{(b)}, \hat{i}_{p_{D_1}}^{(b)}, \hat{m}_{p_{D_1}}^{(b)}, \hat{m}_{D_1}^{\prime(b)}, \hat{k}_{p_{D_1}}^{(b)}, \hat{k}_{D_1}^{\prime(b)}\right)$ such that

    $$\left\{\begin{pmatrix} u_{1p}^n\left(\hat{j}_{p_{D_1}}^{(b)}\right), u_1^n\left(j^{(b)}\middle|\hat{j}_{p_{D_1}}^{(b)}\right), \\ v_1^n\left(\hat{i}_{p_{D_1}}^{(b)}\middle|\hat{j}_{p_{D_1}}^{(b)}, j^{(b)}\right), u_{2p}^n\left(\hat{m}_{p_{D_1}}^{(b)}\middle|\hat{j}_{p_{D_1}}^{(b)}\right), \\ u_2^n\left(m^{(b)}, \hat{m}_{D_1}^{\prime(b)}\middle|\hat{j}_{p_{D_1}}^{(b)}, j^{(b)}, \hat{m}_{p_{D_1}}^{(b)}\right), \\ x_3^n\left(\hat{z}_{p_{D_1}}^{(b)}\middle|\hat{j}_{p_{D_1}}^{(b)}, \hat{m}_{p_{D_1}}^{(b)}\right), \\ v_{12}^n\left(\hat{k}_{p_{D_1}}^{(b)}, \hat{k}_{D_1}^{\prime(b)}\middle|\begin{matrix}\hat{j}_{p_{D_1}}^{(b)}, j^{(b)}, \hat{i}_{p_{D_1}}^{(b)},\\ \hat{m}_{p_{D_1}}^{(b)}, m^{(b)}, \hat{m}_{D_1}^{\prime(b)}\end{matrix}\right), \\ y_1^{n^{(b)}}, \hat{y}_2^n\left(z^{(b)}\middle|\begin{matrix}\hat{j}_{p_{D_1}}^{(b)}, j^{(b)}, \hat{m}_{p_{D_1}}^{(b)},\\ m^{(b)}, \hat{m}_{D_1}^{\prime(b)}, \hat{z}_{p_{D_1}}^{(b)}\end{matrix}\right) \end{pmatrix} \in T_\epsilon^n\right\} \quad \text{(III.11)}$$

    If there is more than one such tuple, decoder 1 chooses one and declares $\left(\hat{j}_{p_{D_1}}^{(b)}, \hat{i}_{D_1}^{(b)}, \hat{m}_{p_{D_1}}^{(b)}, \hat{m}_{D_1}^{\prime(b)}, \hat{k}_{p_{D_1}}^{(b)}, \hat{k}_{D_1}^{\prime(b)}\right)$ was sent. Otherwise, a decoding error is committed. It is proved in Appendix (C) that

    $$\left(\hat{j}_{p_{D_1}}^{(b)}, \hat{i}_{p_{D_1}}^{(b)}, \hat{m}_{p_{D_1}}^{(b)}, \hat{m}_{D_1}^{\prime(b)}, \hat{k}_{p_{D_1}}^{(b)}, \hat{k}_{D_1}^{\prime(b)}\right)$$
    $$\neq \left(j_p^{(b)}, i_p^{(b)}, m_p^{(b)}, m^{*(b)}, k_p^{(b)}, k^{*(b)}\right) \quad \text{(III.12)}$$

    with arbitrarily small probability provided that

    $$\begin{array}{rcll} R_{1p} + R_{11} + L_{2P} + \hat{R} + L_{1B} & \leq & A + B + D & \text{(III.13.A)} \\ R_{11} + L_{2P} + \hat{R} + L_{1B} & \leq & A + B + E & \text{(III.13.B)} \\ R_{11} + \hat{R} + L_{1B} & \leq & A + B + F & \text{(III.13.C)} \\ R_{11} + L_{1B} & \leq & A + G & \text{(III.13.D)} \\ L_{2P} + \hat{R} + L_{1B} & \leq & A + B + H & \text{(III.13.E)} \\ \hat{R} + L_{1B} & \leq & B + I & \text{(III.13.F)} \\ \hat{R} & \leq & I(Y_1; X_3 | U_{1p}, U_1, V_1, U_{2p}, U_2, V_{12}) + B & \text{(III.13.G)} \\ L_{1B} & \leq & J & \text{(III.13.H)} \end{array}$$

    and n is sufficiently large.

**[The Decoding Process of Decoder 2]**

After the transmission of block b, $b = 1, 2, \dots, (B + 2)$, destination 2 receives $y_2^{n^{(b)}}$. The decoding process is divided into two steps as the following:

- The First Decoding Step:

    It is assumed that decoder 2 knows $\left(j_p^{(b)}\right) = \left(j^{(b-1)}\right)$, $\left(m_p^{(b)}\right) = \left(m^{(b-1)}\right)$ and $\left(z_p^{(b)}\right) = \left(z^{(b-1)}\right)$ by decoding block $(b - 1)$. According to the observation $y_2^{n^{(b)}}$, decoder 2 chooses a tuple $\left(\hat{j}_{D_2}^{(b)}, \hat{m}_{D_2}^{(b)}, \hat{m}_{D_2}^{\prime(b)}, \hat{l}_{D_2}^{(b)}, \hat{l}_{D_2}^{\prime(b)}\right)$ such that

    $$\left\{\begin{pmatrix} u_{1p}^n\left(j_p^{(b)}\right), u_1^n\left(\hat{j}_{D_2}^{(b)}\middle|j_p^{(b)}\right), u_{2p}^n\left(m_p^{(b)}\middle|j_p^{(b)}\right), \\ u_2^n\left(\hat{m}_{D_2}^{(b)}, \hat{m}_{D_2}^{\prime(b)}\middle|j_p^{(b)}, \hat{j}_{D_2}^{(b)}, m_p^{(b)}\right), \\ v_2^n\left(\hat{l}_{D_2}^{(b)}, \hat{l}_{D_2}^{\prime(b)}\middle|j_p^{(b)}, \hat{j}_{D_2}^{(b)}, m_p^{(b)}, \hat{m}_{D_2}^{(b)}, \hat{m}_{D_2}^{\prime(b)}\right) \\ , x_3^n\left(z_p^{(b)}\middle|j_p^{(b)}, m_p^{(b)}\right), y_2^{n^{(b)}} \end{pmatrix} \in T_\epsilon^n\right\} \quad \text{(III.14)}$$

    If there is more than one such tuple, decoder 2 chooses one and declares $\left(\hat{j}_{D_2}^{(b)}, \hat{m}_{D_2}^{(b)}, \hat{m}_{D_2}^{\prime(b)}, \hat{l}_{D_2}^{(b)}, \hat{l}_{D_2}^{\prime(b)}\right)$ was sent. Otherwise, a decoding error is committed. It is proved in Appendix (C) that

    $$\left(\hat{j}_{D_2}^{(b)}, \hat{m}_{D_2}^{(b)}, \hat{m}_{D_2}^{\prime(b)}, \hat{l}_{D_2}^{(b)}, \hat{l}_{D_2}^{\prime(b)}\right) \neq \left(j^{(b)}, m^{(b)}, m^{*(b)}, l^{(b)}, l^{*(b)}\right) \quad \text{(III.15)}$$

    with arbitrarily small probability of error provided that

    $$\begin{array}{rcll} R_{1P} + L_{2P} + L_{22} & \leq & K & \text{(III.16.A)} \\ L_{2P} + L_{22} & \leq & L & \text{(III.16.B)} \end{array}$$

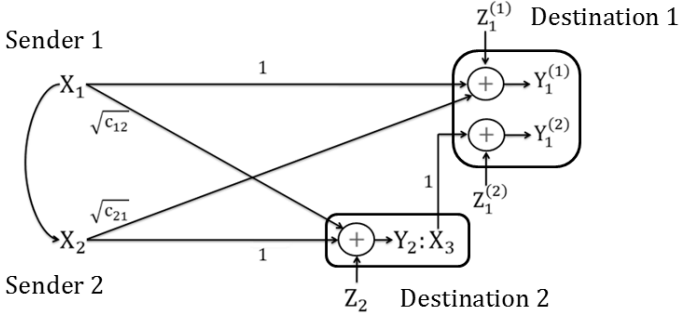

Fig. 2. A special case of the Gaussian CIFC-UDCs.

$$L_{22} \leq M \quad (\text{III.16.C})$$

and n is sufficiently large.

- The Second Decoding Step:

After the first step, $(j^{(b)}, m^{(b)})$ is known to decoder 2. Furthermore, decoder 2 chooses a $\hat{z}_{D2}^{(b)}$ such that

$$\left\{ \begin{array}{c} \left(u_{1p}^n\left(j_p^{(b)}\right), u_1^n\left(j^{(b)}\middle|j_p^{(b)}\right), u_{2p}^n\left(m_p^{(b)}\middle|j_p^{(b)}\right),\right. \\ u_2^n\left(m^{(b)}, m^{*(b)}\middle|j_p^{(b)}, j^{(b)}, m_p^{(b)}\right), \\ x_3^n\left(z_p^{(b)}\middle|j_p^{(b)}, m_p^{(b)}\right), \\ \left.\hat{y}_2^n\left(\hat{z}_{D2}^{(b)}\middle|\begin{array}{c}j_p^{(b)}, j^{(b)}, m_p^{(b)},\\ m^{(b)}, m^{*(b)}, z_p^{(b)}\end{array}\right), y_2^{n^{(b)}}\right) \end{array} \right\} \in T_\epsilon^n \quad (\text{III.17})$$

There will exist such a $\hat{z}_{D2}^{(b)}$ with sufficiently high probability if (III.10) is satisfied, and n goes to infinity. Then decoder 2 sets $z_p^{(b+1)} = z^{(b)} = \hat{z}_{D2}^{(b)}$ and transmits $x_3^n\left(z_p^{(b+1)}\middle|j_p^{(b+1)}, m_p^{(b+1)}\right)$ in block (b+1).

Then we apply *Fourier-Motzkin elimination* [1, pp. 155–156] to eliminate $\hat{R}$, and, after a few steps of manipulations, our region can be obtained.

**Probability of Error Analysis**

This part is provided in Appendix (C). ∎

*Remark 1:* Note that the achievable rate region defined by $\mathcal{R}$ is convex. Henceforth, no convex hull operation or time-sharing is necessary. The proof of the convexity follows the steps in [9, Appendix A] and, therefore, is omitted here.

*Remark 2:* While the unidirectional destination cooperation strategy presented in [42, Theorem 1] only allows destination 2 to forward the public sub-message of PU, $W_{1P}$, in this work, destination 2 is able to forward the public sub-message of CU, $W_{1P}$, on also. This additional cooperation modality is actually an application of IFT.

Here we discuss how our coding scheme performs the interference mitigation via the unidirectional destination cooperation. A special communication setting, depicted in Fig. 2, is used to illustrate our point:

As shown in Fig. 2, destination 1 consists of two component receivers. Notice that the channel link from destination 2 to destination 1 is orthogonal to the rest. Therefore, this channel can be mathematically described as

$$\begin{array}{rcl} Y_1^{(1)} & = & X_1 + \sqrt{c_{21}}X_2 + Z_1^{(1)} \\ Y_1^{(2)} & = & X_3 + Z_1^{(2)} \\ Y_2 & = & X_2 + \sqrt{c_{12}}X_1 + Z_2 \end{array} \quad (\text{III.18})$$

where $X_i$ is the channel input, $i = 1,2,3$, $Z_1^{(1)}$, $Z_1^{(2)}$ and $Z_2$ denote the independently additive Gaussian noises, $Y_1^{(1)}$, $Y_1^{(2)}$ and $Y_2$ denote the channel outputs at the two destinations, and $\sqrt{c_{12}}$ and $\sqrt{c_{21}}$ are the normalized link gains.

For this setting, the secondary destination would forward on part of CU's message to alleviate the interference caused by the transmission of sender 2. In particular, if the power of the noise $Z_1^{(2)}$ is low enough[5], the secondary destination would be able to facilitate the second receiver of destination 1 to decode and understand the complete information about the message $w_2$. Hence, the destination of PU would further know the transmitted symbols of sender 2 by the codebook of CU (i.e., $x_2^n(w_2)$), subtract $\sqrt{c_{21}} \times x_2^n(w_2)$ from $y_1^{(1)n}$ and decode $w_1$ at last. In this manner, CU can communicate without degrading the achievable rate of PU. More interestingly, by combing the idea above with GP precoding, the interferences caused at both the destinations can be alleviated, which is impossible to attain by the strategy of [42, Theorem 1].

## IV. AN OUTER BOUND AND A CAPACITY RESULT FOR THE DISCRETE MEMORYLESS COGNITIVE INTERFERENCE CHANNEL WITH UNIDIRECTIONAL DESTINATION COOPERATION

In this section, we further derive another inner bound and an outer bound for the DM-CIFC-UDC. In general, the two bounds do not meet with each other. But for a class of the DM-CIFC-UDCs, the two bounds are shown to coincide with each other, leading to the capacity result for this special class. At first, we derive the following outer bound.

*Theorem 2:* If $(R_1, R_2)$ lies in the capacity region for the DM-CIFC-UDC,

$$\begin{array}{rcl} R_1 & \leq & I(Y_1; X_1, X_2, X_3) \quad (\text{IV.1.A}) \\ R_2 & \leq & I(Y_2; X_2|X_1, X_3) \quad (\text{IV.1.B}) \\ R_1 + R_2 & \leq & I(Y_1, Y_2; X_1, X_2|X_3) \quad (\text{IV.1.C}) \end{array}$$

taken over the union of all joint distributions $p(x_1, x_2, x_3, y_1, y_2)$.

*Proof:* The proof of this theorem is provided in Appendix (D). ∎

---

[5] As is observed in (III.31), sender 2, destination 2 and the second receiver of destination 1 form a physically degraded relay channel whose capacity coincides with the cut-set bound [2, Chapter 15]. We use the figure below to state this fact. ($S_2$ denotes sender 2, and $D_i$ denotes destination i, i = 1,2.)

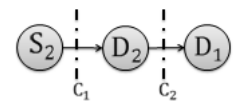

$$C = \min(C_1, C_2)$$

Thus if the power of the noise $Z_1^{(2)}$ is small enough to let $C_1$ to be the bottleneck (i.e., $C_1 \leq C_2$), the secondary destination would be able to inform destination 1 everything "it knows".

The capacity for the DM-CIFC-UDC has been an open problem since its inception. Here we focus on a class of the DM-CIFC-UDCs, the *DM-Z-CIFC-UDC*, whose outer bound derived in theorem 2 is shown to be achievable.

*Definition 5:* The DM-Z-CIFC-UDC is described by a tuple $(\mathcal{X}_1, \mathcal{X}_2, \mathcal{X}_3, \mathcal{Y}_1, \mathcal{Y}_2, p(y_1|x_1,x_3) \times p(y_2|x_1,x_2,x_3))$, where $p(y_1|x_1,x_3)$ and $p(y_2|x_1,x_2,x_3)$ denote the transition probabilities (see Fig. 3).

Furthermore, we specialize our coding scheme in theorem 1 by setting $V_1, U_{2p}, U_2, V_{12}$ and $\widehat{Y}_2$ as $\emptyset$ and letting $X_1 = U_1$, $X_2 = V_2, X_3 = U_{1p}, R_1 = R_{1P}$ and $R_2 = R_{22}$. That is to say, neither sender 1 nor sender 2 performs rate-splitting. However, the message sent from sender 1 is public while the message sent from sender 2 is private. Moreover, we assume the degradedness condition holds:

*Assumption*:
$$p(y_1|y_2,x_1,x_2,x_3) = p(y_1|y_2,x_3) \quad (IV.2)$$
(i.e., $(X_1,X_2) - (Y_2,X_3) - Y_1$)

*Corollary 1:* $\mathcal{P}_a$ denotes the set of all joint probability distributions
$$p(x_1,x_2,x_3,y_1,y_2) \quad (IV.3.A)$$

Let $\mathcal{R}_a(p)$ be the set of all nonnegative rate pairs $(R_1, R_2)$ such that the following inequalities hold:
$$\begin{aligned} R_1 &\leq I(Y_1; X_1, X_3) & (IV.3.B) \\ R_2 &\leq I(Y_2; X_2|X_1, X_3) & (IV.3.C) \\ R_1 + R_2 &\leq I(Y_2; X_1, X_2|X_3) & (IV.3.D) \end{aligned}$$

Then the region $\mathcal{R}_a = \cup_{p(\cdot) \in \mathcal{P}_a} \mathcal{R}_a(p)$ is an achievable rate region for the DM-CIFC-UDC.

*Theorem 3:* The capacity region for the DM-Z-CIFC-UDC with the degradedness condition defined in (IV.2) is given by corollary 1.

*Proof:* In the DM-Z-CIFC-UDC, the following Markov chain holds:
$$X_2 - (X_1, X_3) - Y_1 \quad (IV.4)$$

Thus, the right hand side of the inequality, $I(Y_1; X_1, X_2, X_3)$, in (IV.1.A) can be rewritten as $I(Y_1; X_1, X_3)$. For (IV.1.C),
$$\begin{aligned} R_1 + R_2 &\leq I(Y_1, Y_2; X_1, X_2|X_3) \\ &= I(Y_2; X_1, X_2|X_3) + I(Y_1; X_1, X_2|Y_2, X_3) \end{aligned} \quad (IV.5)$$

The second term in (IV.5) is zero due to (IV.2). Hence, the inner and outer bounds coincide with each other, and the capacity result for this class is obtained. ∎

## V. COMPARISON WITH OTHER EXISTING ACHIEVABLE RATE REGIONS

In this section, we show that our achievable rate region derived in theorem 1 subsumes:

- Chu's achievable rate region for the DM-CIFC-UDC [42, Theorem 1], denoted as $\mathcal{R}_C$.

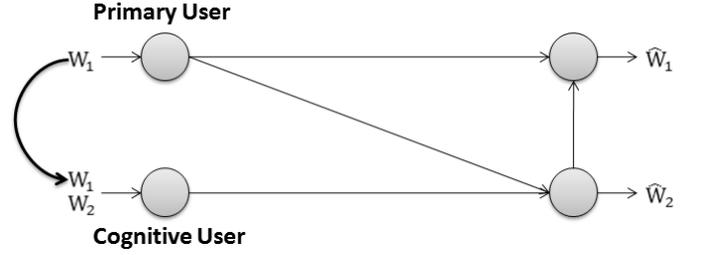

Fig. 3. The DM-Z-CIFC-UDC.

- The largest known achievable rate region, by Rini et al., for the CIFC [47, Theorem V.1], denoted as $\mathcal{R}_{RTD}$.
- The largest known achievable rate region, by Bross, for the partially cooperative relay broadcast channel [38, Theorem 2], denoted as $\mathcal{R}_B$. Besides, a potentially larger region is derived.
- Marton region [4, Theorem 2].
- The largest known achievable rate for the relay channel, CMG rate [15].

### A. Chu's region for the CIFC-UDC

Chu's region is derived by a combination of Gel'fand-Pinsker binning and decode-and-forward relaying strategy which is also integrated into our coding scheme. Notice that the right hand sides of the inequalities (4) and (6) in [42] are incorrect and should be replaced by $I(S; W, Y_1|V, T, Q)$ and $I(T, W, V, S; Y_1|Q)$ respectively.

To prove the inclusion, firstly we utilize theorem 1 to provide a region $\mathcal{R}_{sub} \subseteq \mathcal{R}$ and prove that $\mathcal{R}_{sub}$ subsumes $\mathcal{R}_C$.

*Corollary 2:* Theorem 1 includes $\mathcal{R}_C$.

*Proof:* According to theorem 1, by setting $U_{2p}, V_{12}$ and $\widehat{Y}_2$ as $\emptyset$, $U_{1p}$ to be equal to $X_3$ and $V_1$ to be equal to $X_1$, the following region $\mathcal{R}_{sub}$ is certainly achievable:

$$\begin{aligned} R_{1p} + R_{11} + R_{2P} &\leq I(Y_1; X_3, U_1, X_1, U_2) & (V.1.A) \\ R_{11} + R_{2P} &\leq I(Y_1; X_1, U_2|X_3, U_1) & (V.1.B) \\ R_{2P} &\leq I(Y_1; U_2|X_3, U_1, X_1) & (V.1.C) \\ R_{11} &\leq I(Y_1, U_2; X_1|X_3, U_1) & (V.1.D) \\ R_{1P} + R_{2P} + R_{22} &\leq I(Y_2; U_1, U_2, V_2|X_3) \\ &\quad - I(X_1; U_2, V_2|X_3, U_1) & (V.1.E) \\ R_{2P} + R_{22} &\leq I(Y_2; U_2, V_2|X_3, U_1) \\ &\quad - I(X_1; U_2, V_2|X_3, U_1) & (V.1.F) \\ R_{22} &\leq I(Y_2; V_2|X_3, U_1, U_2) \\ &\quad - I(X_1; V_2|X_3, U_1, U_2) & (V.1.G) \end{aligned}$$

taken over the union of all distributions
$$\begin{aligned} &p(x_3)p(u_1|x_3)p(x_1|x_3,u_1)p(u_2|x_3,u_1,x_1) \\ &\times p(v_2|x_3,u_1,x_1)p(x_2|x_3,u_1,x_1,u_2,v_2) \\ &\times p(y_1,y_2|x_1,x_2,x_3) \end{aligned} \quad (V.1.H)$$

∎

Notice that with this particular factorization, $U_2 - (X_3, U_1, X_1) - V_2$ forms a Markov chain. Thus,
$$I(X_1; V_2|X_3, U_1, U_2) = I(X_1; V_2|X_3, U_1) - I(U_2; V_2|X_3, U_1) \quad (V.2)$$
and

$$I(X_1; U_2, V_2|X_3, U_1) = I(X_1; U_2|X_3, U_1) + I(X_1; V_2|X_3, U_1)$$
$$-I(U_2; V_2|X_3, U_1) \quad \text{(V.3)}$$

By substituting (V.2) for $I(X_1; V_2|X_3, U_1, U_2)$ in (V.1.G), (V.3) for $I(X_1; U_2, V_2|X_3, U_1)$ in (V.1.E) and (V.1.F), and, with some manipulations, (V.1.A) − (V.1.G) can be rewritten as

$$R_{1P} + R_{11} + R_{2P} \le I(Y_1; X_3, U_1, X_1, U_2) \quad \text{(V.4.A)}$$
$$R_{11} + R_{2P} \le I(Y_1, U_2; X_1|X_3, U_1) + I(Y_1; U_2|X_3)$$
$$-I(U_1, X_1; U_2|X_3) + I(U_1; U_2|X_3, Y_1) \quad \text{(V.4.B)}$$
$$R_{2P} \le I(Y_1, U_1, X_1; U_2|X_3) - I(U_1, X_1; U_2|X_3) \quad \text{(V.4.C)}$$
$$R_{11} \le I(Y_1, U_2; X_1|X_3, U_1) \quad \text{(V.4.D)}$$
$$R_{1P} + R_{2P} + R_{22} \le I(Y_2; U_1, U_2, V_2|X_3) + I(U_2; V_2|X_3)$$
$$+I(U_1; U_2, V_2|X_3) - I(U_1, X_1; V_2|X_3) - I(U_1, X_1; U_2|X_3) \quad \text{(V.4.E)}$$
$$R_{2P} + R_{22} \le I(Y_2; U_2, V_2|X_3, U_1) - I(X_1; U_2|X_3, U_1)$$
$$-I(X_1; V_2|X_3, U_1) + I(U_2; V_2|X_3, U_1) \quad \text{(V.4.F)}$$
$$R_{22} \le I(Y_2, U_1, U_2; V_2|X_3) - I(U_1, X_1; V_2|X_3) \quad \text{(V.4.G)}$$

taken over the union of all distributions

$$p(x_3)p(u_1|x_3)p(x_1|x_3, u_1)p(u_2|x_3, u_1, x_1)$$
$$\times p(v_2|x_3, u_1, x_1)p(x_2|x_3, u_1, x_1, u_2, v_2) \quad \text{(V.4.H)}$$
$$\times p(y_1, y_2|x_1, x_2, x_3)$$

Then let us modify $\mathcal{R}_c$. Firstly, since T is decoded at both the decoders, the time sharing random Q can be incorporated with T. Hence, the time-sharing random variable Q can be dropped. In addition, following the argument of [11, Appendix D], we can show that, without loss of generality, we can set $X_1$ and $X_3$ to be the deterministic functions (i.e., $X_1$ as a function of $(T, V, S)$ and $X_3$ as a function of $(T)$) and insert them into the mutual information. With these considerations, we can express $\mathcal{R}_c$ as:

$$R_{1R} + R_{11} + R_{2P} \le I(T, X_3, W, V, S, X_1; Y_1) \quad \text{(V.5.A)}$$
$$R_{11} + R_{2P} \le I(S, X_1; W, Y_1|V, T, X_3) + I(W; Y_1|T, X_3)$$
$$-I(V, S, X_1; W|T, X_3) \quad \text{(V.5.B)}$$
$$R_{11} \le I(S, X_1; W, Y_1|V, T, X_3) \quad \text{(V.5.C)}$$
$$R_{2P} \le I(W; Y_1|T, X_3) - I(V, S, X_1; W|T, X_3) \quad \text{(V.5.D)}$$
$$R_{1R} + R_{22} + R_{2P} \le I(V, U, W; Y_2|T, X_3) + I(V; U, W|T, X_3)$$
$$+I(W; U|T, X_3) - I(V, S, X_1; U|T, X_3) - I(V, S, X_1; W|T, X_3) \quad \text{(V.5.E)}$$
$$R_{1R} + R_{22} \le I(V, U; W, Y_2|T, X_3) + I(U; V|T, X_3)$$
$$-I(V, S, X_1; U|T, X_3) \quad \text{(V.5.F)}$$
$$R_{1R} + R_{2P} \le I(V, W; U, Y_2|T, X_3) + I(W; V|T, X_3)$$
$$-I(V, S, X_1; W|T, X_3) \quad \text{(V.5.G)}$$
$$R_{22} + R_{2P} \le I(U, W; V, Y_2|T, X_3) + I(U; W|T, X_3)$$
$$-I(V, S, X_1; U|T, X_3) - I(V, S, X_1; W|T, X_3) \quad \text{(V.5.H)}$$
$$R_{1R} \le I(V; U, W, Y_2|T, X_3) \quad \text{(V.5.I)}$$
$$R_{22} \le I(U; V, W, Y_2|T, X_3) - I(V, S, X_1; U|T, X_3) \quad \text{(V.5.J)}$$
$$R_{2P} \le I(W; V, U, Y_2|T, X_3) - I(V, S, X_1; W|T, X_3) \quad \text{(V.5.K)}$$

taken over the union of all distributions

$$p(t, x_3)p(v|t, x_3)p(s|t, x_3, v)p(x_1|t, x_3, v, s)$$
$$\times p(w|t, x_3, v, s, x_1)p(u|t, x_3, v, s, x_1) \quad \text{(V.5.L)}$$
$$\times p(x_2|t, x_3, v, s, x_1, w, u)p(y_1, y_2|s, x_1, x_2, t, x_3)$$

Notice that (V.5.B) is obtained simply by adding (4) and (5) together in [42]. Trivially, this operation has no impact on the region but makes our proof much easier. In addition, one random variable can be eliminated by noticing (V.5.L)

$$p(t, x_3)p(v|t, x_3)p(s|t, x_3, v)p(x_1|t, x_3, v, s)p(w|t, x_3, v, s, x_1)$$
$$\times p(u|t, x_3, v, s, x_1)p(x_2|t, x_3, v, s, x_1, w, u)p(y_1, y_2|s, x_1, x_2, t, x_3)$$
$$= p(t, x_3)p(v|t, x_3)p(s, x_1|t, x_3, v)p(w|t, x_3, v, s, x_1)$$
$$\times p(u|t, x_3, v, s, x_1)p(x_2|t, x_3, v, s, x_1, w, u)p(y_1, y_2|s, x_1, x_2, t, x_3)$$

$$\times p(y_1, y_2|s, x_1, x_2, t, x_3) \quad \text{(V.6)}$$

and setting $[S, X_1] = S'$ and $[T, X_3] = T'$, we obtain the region

$$R_{1R} + R_{11} + R_{2P} \le I(T', W, V, S'; Y_1) \quad \text{(V.7.A)}$$
$$R_{11} + R_{2P} \le I(S'; W, Y_1|V, T') + I(W; Y_1|T')$$
$$-I(V, S'; W|T') \quad \text{(V.7.B)}$$
$$R_{11} \le I(S'; W, Y_1|V, T') \quad \text{(V.7.C)}$$
$$R_{2P} \le I(W; Y_1|T') - I(V, S'; W|T') \quad \text{(V.7.D)}$$
$$R_{1R} + R_{22} + R_{2P} \le I(V, U, W; Y_2|T') + I(V; U, W|T') + I(W; U|T')$$
$$-I(V, S'; U|T') - I(V, S'; W|T') \quad \text{(V.7.E)}$$
$$R_{1R} + R_{22} \le I(V, U; W, Y_2|T') + I(U; V|T') - I(V, S'; U|T') \quad \text{(V.7.F)}$$
$$R_{1R} + R_{2P} \le I(V, W; U, Y_2|T')$$
$$+I(W; V|T') - I(V, S'; W|T') \quad \text{(V.7.G)}$$
$$R_{22} + R_{2P} \le I(U, W; V, Y_2|T') + I(U; W|T')$$
$$-I(V, S'; U|T') - I(V, S'; W|T') \quad \text{(V.7.H)}$$
$$R_{1R} \le I(V; U, W, Y_2|T') \quad \text{(V.7.I)}$$
$$R_{22} \le I(U; V, W, Y_2|T') - I(V, S'; U|T') \quad \text{(V.7.J)}$$
$$R_{2P} \le I(W; V, U, Y_2|T') - I(V, S'; W|T') \quad \text{(V.7.K)}$$

taken over the union of all distributions

$$p(t')p(v|t')p(s'|v, t')p(w|s', v, t')p(u|s', v, t')$$
$$\times p(x_2|u, w, s', v, t')p(y_1, y_2|s', x_2, t') \quad \text{(V.7.L)}$$

Lastly, by comparing (V.4) with (V.7) term by term, it is not difficult to observe that (V.7) introduces more rate constraints than (V.4). Hence, the inclusion follows. ∎

### B. Rini et al.,'s region for the CIFC

From an information-theoretic perspective, the CIFC-UDC includes the CIFC as a special case. More specifically, by disabling the destination cooperation, the CIFC-UDC is equivalent to the CIFC in all aspects. Hence, theorem 1 can be utilized to derive an achievable rate region for the CIFC. More importantly, in this part, we show that our region subsumes the largest known achievable rate region for the CIFC, $\mathcal{R}_{RTD}$.

*Corollary 3:* Theorem 1 includes $\mathcal{R}_{RTD}$.

*Proof:* by setting the random variables associated with the unidirectional destination cooperation in theorem 1, $U_{1p}$, $U_{2p}$, $X_3$ and $\widehat{Y}_2$ as ∅, our theorem is equivalent to $\mathcal{R}_{RTD}$, and the inclusion follows. ∎

### C. Bross' region for the partially cooperative relay broadcast channel and Marton region

At first, an achievable rate region for the PCRBC, $\mathcal{R}_{PCRBC}$, is derived from and subsumed by theorem 1. Then, we demonstrate that this region $\mathcal{R}_{PCRBC}$ is potentially larger than $\mathcal{R}_B$ by proving the inclusion $\mathcal{R}_B \subseteq \mathcal{R}_{PCRBC}$. As a result, the inclusion of Marton region follows (see *Remark 2* in [31]).

*Corollary 4:* $\mathcal{P}_b$ denotes the set of all joint probability distributions

$$p(u_{2p}, u_2, v_{12}, v_2, x_2)p(x_3|u_{2p})$$
$$\times p(y_1, y_2|x_1, x_2, x_3)p(\hat{y}_2|u_{2p}, u_2, x_3, y_2) \quad \text{(V.8.A)}$$

Let $\mathcal{R}_b(p)$ be the set of all nonnegative rate triple $(R_{2P}, R_{22}, R_{1B})$ such that the following constraints hold:

$$R'_{1B} + R'_{22} \geq I(V_{12}; V_2 | U_{2p}, U_2) \quad \text{(V.8.B)}$$
$$R_{2P} + (R_{1B} + R'_{1B}) \leq I(\widehat{Y}_2; V_{12}, Y_1 | U_{2p}, U_2, X_3)$$
$$-I(Y_2; \widehat{Y}_2 | U_{2p}, U_2, X_3) + I(Y_1; U_{2p}, U_2, V_{12}, X_3) \quad \text{(V.8.C)}$$

$$(R_{1B} + R'_{1B}) \leq \min \begin{pmatrix} \left(I(Y_1, \widehat{Y}_2; V_{12} | U_{2p}, U_2, X_3)\right), \\ \begin{pmatrix} I(Y_1; V_{12}, X_3 | U_{2p}, U_2) \\ +I(\widehat{Y}_2; V_{12}, Y_1 | U_{2p}, U_2, X_3) \\ -I(Y_2; \widehat{Y}_2 | U_{2p}, U_2, X_3) \end{pmatrix} \end{pmatrix} \quad \text{(V.8.D)}$$

$$R_{2P} + (R_{22} + R'_{22}) \leq I(Y_2; U_2, V_2 | U_{2p}, X_3) \quad \text{(V.8.E)}$$
$$(R_{22} + R'_{22}) \leq I(Y_2; V_2 | U_{2p}, U_2, X_3) \quad \text{(V.8.F)}$$
$$I(Y_2; \widehat{Y}_2 | U_{2p}, U_2, X_3) \leq I(Y_1; X_3 | U_{2p}, U_2, V_{12})$$
$$+I(\widehat{Y}_2; V_{12}, Y_1 | U_{2p}, U_2, X_3) \quad \text{(V.8.G)}$$

Then the region $\mathcal{R}_{\mathcal{PCRBC}} = \cup_{p(\cdot) \in \mathcal{P}_b} \mathcal{R}_b(p)$ is an achievable rate region for the discrete memoryless PCRBC.

*Proof:* By setting $U_{1p}, U_1, V_1$ and $X_1$ as $\emptyset$ in theorem 1, this corollary is proved. ∎

It remains difficult to compare (V.8) with $\mathcal{R}_{\mathcal{B}}$. At first, we shrink this region $\mathcal{R}_{\mathcal{PCRBC}}$ by replacing (V.8.G) with a tighter bound:

*Corollary 5:* $\mathcal{R}_{\mathcal{B}} \subseteq \mathcal{R}_{\mathcal{PCRBC}} \subseteq \mathcal{R}$.

*Proof:* The left hand side of the inequality in (V.8.G):

$$I(Y_2; \widehat{Y}_2 | U_{2p}, U_2, X_3)$$
$$= I(Y_1, Y_2; \widehat{Y}_2 | U_{2p}, U_2, X_3)$$
$$= I(Y_1; \widehat{Y}_2 | U_{2p}, U_2, X_3) + I(Y_2; \widehat{Y}_2 | U_{2p}, U_2, X_3, Y_1) \quad \text{(V.9)}$$

where the first equality follows from the Markov chain,

$$Y_1 - (U_{2p}, U_2, X_3, Y_2) - \widehat{Y}_2 \quad \text{(V.10)}$$

For the terms on the right hand side of the inequality in (V.8.G):

$$I(Y_1; X_3 | U_{2p}, U_2, V_{12})$$
$$= H(X_3 | U_{2p}, U_2, V_{12}) - H(X_3 | U_{2p}, U_2, V_{12}, Y_1)$$
$$= H(X_3 | U_{2p}, U_2) - H(X_3 | U_{2p}, U_2, V_{12}, Y_1)$$
$$\geq H(X_3 | U_{2p}, U_2) - H(X_3 | U_{2p}, U_2, Y_1)$$
$$= I(Y_1; X_3 | U_{2p}, U_2) \quad \text{(V.11)}$$

$$I(\widehat{Y}_2; V_{12}, Y_1 | U_{2p}, U_2, X_3)$$
$$= I(\widehat{Y}_2; Y_1 | U_{2p}, U_2, X_3) + I(\widehat{Y}_2; V_{12} | U_{2p}, U_2, X_3, Y_1)$$
$$\geq I(\widehat{Y}_2; Y_1 | U_{2p}, U_2, X_3) \quad \text{(V.12)}$$

By substituting (V.9), (V.11) and (V.12) for the terms in (V.8.G), we obtain an tighter bound (than (V.8.G)):

$$I(Y_2; \widehat{Y}_2 | U_{2p}, U_2, X_3, Y_1) \leq I(Y_1; X_3 | U_{2p}, U_2) \quad \text{(V.13)}$$

Thus, the region defined by (V.8.A) − (V.8.F) and (V.13) is also achievable for the discrete memoryless PCRBC. Then let us perform some modifications on this region.

The second term in the parenthesis of (V.8.D) can be dropped due to this fact:

$$I(Y_1; V_{12}, X_3 | U_{2p}, U_2) + I(\widehat{Y}_2; V_{12}, Y_1 | U_{2p}, U_2, X_3)$$

$$-I(Y_2; \widehat{Y}_2 | U_{2p}, U_2, X_3)$$
$$= \left(I(Y_1; X_3 | U_{2p}, U_2) + I(Y_1; V_{12} | U_{2p}, U_2, X_3)\right)$$
$$\quad +I(\widehat{Y}_2; V_{12}, Y_1 | U_{2p}, U_2, X_3) - I(Y_2; \widehat{Y}_2 | U_{2p}, U_2, X_3)$$
$$= \left(I(Y_1; X_3 | U_{2p}, U_2) + I(\widehat{Y}_2; Y_1 | U_{2p}, U_2, X_3)\right)$$
$$\quad + \left(I(Y_1; V_{12} | U_{2p}, U_2, X_3) + I(\widehat{Y}_2; V_{12} | U_{2p}, U_2, X_3, Y_1)\right)$$
$$\quad -I(Y_2; \widehat{Y}_2 | U_{2p}, U_2, X_3)$$
$$= I(Y_1, \widehat{Y}_2; V_{12} | U_{2p}, U_2, X_3) + I(X_3, \widehat{Y}_2; Y_1 | U_{2p}, U_2)$$
$$\quad - \left(I(Y_2; \widehat{Y}_2 | U_{2p}, U_2, X_3)\right)$$
$$= I(Y_1, \widehat{Y}_2; V_{12} | U_{2p}, U_2, X_3) + I(X_3, \widehat{Y}_2; Y_1 | U_{2p}, U_2)$$
$$\quad - \left(I(Y_1, Y_2; \widehat{Y}_2 | U_{2p}, U_2, X_3)\right) \quad \text{(V.14)}$$
$$= I(Y_1, \widehat{Y}_2; V_{12} | U_{2p}, U_2, X_3) + \left(I(X_3, \widehat{Y}_2; Y_1 | U_{2p}, U_2)\right)$$
$$\quad -I(Y_1; \widehat{Y}_2 | U_{2p}, U_2, X_3) - I(Y_2; \widehat{Y}_2 | U_{2p}, U_2, X_3, Y_1)$$
$$= I(Y_1, \widehat{Y}_2; V_{12} | U_{2p}, U_2, X_3)$$
$$\quad + \left(I(X_3; Y_1 | U_{2p}, U_2) + I(\widehat{Y}_2; Y_1 | U_{2p}, U_2, X_3)\right)$$
$$\quad -I(Y_1; \widehat{Y}_2 | U_{2p}, U_2, X_3) - I(Y_2; \widehat{Y}_2 | U_{2p}, U_2, X_3, Y_1)$$
$$= I(Y_1, \widehat{Y}_2; V_{12} | U_{2p}, U_2, X_3)$$
$$\quad + \left(I(X_3; Y_1 | U_{2p}, U_2) - I(Y_2; \widehat{Y}_2 | U_{2p}, U_2, X_3, Y_1)\right)$$
$$\geq I(Y_1, \widehat{Y}_2; V_{12} | U_{2p}, U_2, X_3) \quad \text{(V.15)}$$

where (V.14) follows from (V.10), and (V.15) follows from (V.13). Thus, we obtain the following region:

$$R'_{1B} + R'_{22} \geq I(V_{12}; V_2 | U_{2p}, U_2) \quad \text{(V.16.A)}$$
$$R_{2P} + (R_{1B} + R'_{1B}) \leq I(\widehat{Y}_2; V_{12}, Y_1 | U_{2p}, U_2, X_3)$$
$$-I(Y_2; \widehat{Y}_2 | U_{2p}, U_2, X_3) + I(Y_1; U_{2p}, U_2, V_{12}, X_3) \quad \text{(V.16.B)}$$
$$(R_{1B} + R'_{1B}) \leq I(Y_1, \widehat{Y}_2; V_{12} | U_{2p}, U_2, X_3) \quad \text{(V.16.C)}$$
$$R_{2P} + (R_{22} + R'_{22}) \leq I(Y_2; U_2, V_2 | U_{2p}, X_3) \quad \text{(V.16.D)}$$
$$(R_{22} + R'_{22}) \leq I(Y_2; V_2 | U_{2p}, U_2, X_3) \quad \text{(V.16.E)}$$
$$I(Y_2; \widehat{Y}_2 | U_{2p}, U_2, X_3, Y_1) \leq I(Y_1; X_3 | U_{2p}, U_2) \quad \text{(V.16.F)}$$

taken over the union of all distributions

$$p(u_{2p}, u_2, v_{12}, v_2, x_2) p(x_3 | u_{2p}) \\ \times p(y_1, y_2 | x_1, x_2, x_3) p(\widehat{y}_2 | u_{2p}, u_2, x_3, y_2) \quad \text{(V.16.G)}$$

This region does not impose an explicit constraint on the common message $R_{2P}$. We further shrink the region (V.16) by adding the constraint below:

$$R_{2P} \leq I(Y_1; U_{2p}, U_2, X_3, \widehat{Y}_2) - I(Y_2; \widehat{Y}_2 | U_{2p}, U_2, X_3) \quad \text{(V.17)}$$

, and, according to (V.16.B), for the most restrictive situation (i.e., $R_{2P}$ is chosen to coincide with its upper-bound), $(R_{1B} + R'_{1B})$ should lie in the following range:

$$(R_{1B} + R'_{1B}) \leq I(\widehat{Y}_2; V_{12}, Y_1 | U_{2p}, U_2, X_3)$$
$$\quad -I(Y_2; \widehat{Y}_2 | U_{2p}, U_2, X_3) + I(Y_1; U_{2p}, U_2, V_{12}, X_3)$$
$$\quad - \left(I(Y_1; U_{2p}, U_2, X_3, \widehat{Y}_2) - I(Y_2; \widehat{Y}_2 | U_{2p}, U_2, X_3)\right)$$
$$= I(\widehat{Y}_2; V_{12} | U_{2p}, U_2, X_3) + I(\widehat{Y}_2; Y_1 | U_{2p}, U_2, X_3, V_{12})$$
$$\quad -I(Y_2; \widehat{Y}_2 | U_{2p}, U_2, X_3) + I(Y_1; U_{2p}, U_2, V_{12}, X_3)$$
$$\quad - \left(I(Y_1; U_{2p}, U_2, X_3, \widehat{Y}_2) - I(Y_2; \widehat{Y}_2 | U_{2p}, U_2, X_3)\right)$$
$$= I(\widehat{Y}_2; V_{12} | U_{2p}, U_2, X_3) + I(\widehat{Y}_2; Y_1 | U_{2p}, U_2, X_3, V_{12}; Y_1)$$
$$\quad -I(Y_1; U_{2p}, U_2, X_3, \widehat{Y}_2)$$
$$= I(\widehat{Y}_2; V_{12} | U_{2p}, U_2, X_3) + I(V_{12}; Y_1 | U_{2p}, U_2, X_3, \widehat{Y}_2)$$
$$= I(V_{12}; Y_1, \widehat{Y}_2 | U_{2p}, U_2, X_3) \quad \text{(V.18)}$$

which is exactly the same as (V.16.C) and impose no additional constraint. To sum up, the following region is achievable for the discrete memoryless PCRBC and subsumed in $\mathcal{R}_{PCRBC}$:

$$R'_{1B} + R'_{22} \geq I(V_{12}; V_2 | U_{2p}, U_2) \quad (V.19.A)$$
$$R_{2P} \leq I(Y_1; U_{2p}, U_2, X_3, \widehat{Y}_2) - I(Y_2; \widehat{Y}_2 | U_{2p}, U_2, X_3) \quad (V.19.B)$$
$$(R_{1B} + R'_{1B}) \leq I(Y_1, \widehat{Y}_2; V_{12} | U_{2p}, U_2, X_3) \quad (V.19.C)$$
$$R_{2P} + (R_{22} + R'_{22}) \leq I(Y_2; U_2, V_2 | U_{2p}, X_3) \quad (V.19.D)$$
$$(R_{22} + R'_{22}) \leq I(Y_2; V_2 | U_{2p}, U_2, X_3) \quad (V.19.E)$$
$$I(Y_2; \widehat{Y}_2 | U_{2p}, U_2, X_3, Y_1) \leq I(Y_1; X_3 | U_{2p}, U_2) \quad (V.19.F)$$

taken over the union of all distributions

$$p(u_{2p}, u_2, v_{12}, v_2, x_2)p(x_3|u_{2p})$$
$$\times p(y_1, y_2 | x_1, x_2, x_3)p(\widehat{y}_2 | u_{2p}, u_2, x_3, y_2) \quad (V.19.G)$$

After scrutinizing the region (V.19) above, one can find out that this is actually equivalent to $\mathcal{R}_B$, Bross' region, and the inclusion follows. ∎

### D. CMG Rate

*Corollary 6:* CMG rate is included in theorem 1.

*Proof:* By setting $U_{2p}, U_2, V_{12}, V_2$ and $X_2$ as $\emptyset$ in theorem 1, this claim follows. ∎

## VI. CONCLUSION

In this paper, a new unified achievable rate region for the DM-CIFC-UDC is derived. This region is further proved to include the largest known achievable rate regions for the broadcast channel, the relay channel, the cognitive radio interference channel and the partially cooperative relay broadcast channel. More importantly, a potentially larger achievable rate region for the partially cooperative relay broadcast channel is derived.

In addition, how to assist the decoder of PU to alleviate the interference caused by the transmission of CU via the unidirectional destination cooperation is also investigated. As a matter of fact, this interference mitigation strategy can be considered as a complementary part of the non-causal cognition originally proposed by Devroye et al. [12], [16] which demonstrated that by Gel'fand-Pinsker precoding, the interference caused by PU on the secondary destination can be mitigated.

An outer bound of the capacity region for the CIFC-UDC is also derived in this paper. Although this outer bound is not tight in general, for a class of the DM-CIFC-UDCs, our achievable rate region meets with this outer bound, resulting in the characterization of the capacity region for this class.

## APPENDIX (A)
### PROOF OF LEMMA 1

*Lemma 1:* For fixed $(j_p, j, i_p, m_p, m)$, with sufficiently high probability, encoder 2 can find at least one $m'$ such that

$$\left\{ \begin{pmatrix} u_{1p}^n(j_p), u_1^n(j|j_p), v_1^n(i_p|j_p, j), \\ u_{2p}^n(m_p|j_p), u_2^n(m, m'|j_p, j, m_p) \end{pmatrix} \in T_\epsilon^n \right\} \quad (A1)$$

provided that

$$R'_{2P} \geq I(V_1; U_2 | U_{1p}, U_1, U_{2p}) \quad (A2)$$

and n is sufficiently large.

*Proof:* This lemma can be proved by the similar approach used in the rate-distortion problem [2, Chapter 10]. Since it is assumed that all the sub-message tuples are equiprobable, without losing generality, we assume that $(j_p, j, i_p, m_p, m) = (1,1,1,1,1)$.

$$\Pr \left\{ \bigcap_{m'=1}^{2^{nR'_{2P}}} \left\{ \begin{pmatrix} u_{1p}^n(1), u_1^n(1|1), v_1^n(1|1,1), \\ u_{2p}^n(1|1), u_2^n(1, m'|1,1,1) \end{pmatrix} \notin T_\epsilon^n \right\} \right\}$$
$$= \left( 1 - \Pr\{(u_{1p}^n, u_1^n, v_1^n, u_{2p}^n, u_2^n) \in T_\epsilon^n\} \right)^{2^{nR'_{2P}}} \quad (A3)$$

Furthermore, the latter term in the parenthesis of (A3) can be lower-bounded as

$$\Pr\{(u_{1p}^n, u_1^n, v_1^n, u_{2p}^n, u_2^n) \in T_\epsilon^n\}$$
$$= \sum_{u_2^n \in T_\epsilon^n} p(u_2^n | u_{1p}^n, u_1^n, u_{2p}^n)$$
$$\geq 2^{n(H(U_2|U_{1p}, U_1, V_1, U_{2p}) - \epsilon)} 2^{-n(H(U_2|U_{1p}, U_1, U_{2p}) + \epsilon)}$$
$$= 2^{-n(I(U_2; V_1 | U_{1p}, U_1, U_{2p}) + 2\epsilon)} \quad (A4)$$

In addition, by employing the inequality, $(1 - x)^m \leq e^{-mx}$, (A3) can be upper-bounded as

$$\Pr \left\{ \bigcap_{m'=1}^{2^{nR'_{2P}}} \left\{ \begin{pmatrix} u_{1p}^n(1), u_1^n(1|1), v_1^n(1|1,1), \\ u_{2p}^n(1|1), u_2^n(1, m'|1,1,1) \end{pmatrix} \notin T_\epsilon^n \right\} \right\}$$
$$\leq \left( 1 - 2^{-n(I(U_2; V_1 | U_{1p}, U_1, U_{2p}) + 2\epsilon)} \right)^{2^{nR'_{2P}}}$$
$$\leq \exp\left( -\left( 2^{-n(I(U_2; V_1 | U_{1p}, U_1, U_{2p}) + 2\epsilon)} \times 2^{nR'_{2P}} \right) \right)$$
$$= \exp\left( -2^{n(R'_{2P} - I(U_2; V_1 | U_{1p}, U_1, U_{2p}) - 2\epsilon)} \right) \quad (A5)$$

Hence, according to (A5), the probability (A3), can be made as small as possible provided that (A2) is satisfied, and n goes to infinity. ∎

## APPENDIX (B)
### PROOF OF LEMMA 2

*Lemma 2:* For fixed $(j_p, j, i_p, m_p, m, k_p, l)$ and previously found $m^*$, with sufficiently high probability, encoder 2 can find at least one pair $(k', l')$ such that

$$\left\{ \begin{pmatrix} u_{1p}^n(j_p), u_1^n(j|j_p), v_1^n(i_p|j_p, j), \\ u_{2p}^n(m_p|j_p), u_2^n(m, m^*|j_p, j, m_p), \\ v_{12}^n(k_p, k'|j_p, j, i_p, m_p, m, m^*), \\ v_2^n(l, l'|j_p, j, m_p, m, m^*) \end{pmatrix} \in T_\epsilon^n \right\} \quad (B1)$$

provided that

$$R'_{1B} \geq 0 \quad (B2.A)$$
$$R'_{22} \geq I(V_1; V_2 | U_{1p}, U_1, U_{2p}, U_2) \quad (B2.B)$$
$$R'_{1B} + R'_{22} \geq I(V_1, V_{12}; V_2 | U_{1p}, U_1, U_{2p}, U_2) \quad (B2.C)$$

and n is sufficiently large.

*Proof:* Lemma 2 can be considered as an extension of Marton binning. Since it is assumed that all the sub-message tuples are equiprobable, without loss of generality, we assume that $(j_p, j, i_p, m_p, m, k_p, l) = (1,1,1,1,1,1,1)$.

$$\Pr\left\{\bigcap_{(k',l')}\left\{\begin{pmatrix}u_{1p}^n(1),u_1^n(1|1),v_1^n(1|1,1,1),\\u_{2p}^n(1|1),u_2^n(1,m^*|1,1,1),\\v_{12}^n(1,k'|1,1,1,1,1,m^*),\\v_2^n(1,l'|1,1,1,1,m^*)\end{pmatrix}\notin T_\epsilon^n\right\}\right\} \quad \text{(B3)}$$

The probability above involves an intersection of dependent events, rather than a union. We employ the approach in [8] to deal with this problem. Let $I(k',l')$ be the indicator function where the event

$$\left\{\begin{pmatrix}u_{1p}^n(1),u_1^n(1|1),v_1^n(1|1,1,1),\\u_{2p}^n(1|1),u_2^n(1,m^*|1,1,1),\\v_{12}^n(1,k'|1,1,1,1,1,m^*),\\v_2^n(1,l'|1,1,1,1,m^*)\end{pmatrix}\in T_\epsilon^n\right\} \quad \text{(B4)}$$

occurs. Furthermore, we define a random variable as

$$K=\sum_{k'=1}^{2^{nR'_{1B}}}\sum_{l'=1}^{2^{nR'_{22}}}I(k',l') \quad \text{(B5)}$$

Equation (B3) can be bounded as

$$\Pr\left\{\bigcap_{(k',l')}\left\{\begin{pmatrix}u_{1p}^n(1),u_1^n(1|1),v_1^n(1|1,1,1),\\u_{2p}^n(1|1),u_2^n(1,m^*|1,1,1),\\v_{12}^n(1,k'|1,1,1,1,1,m^*),\\v_2^n(1,l'|1,1,1,1,m^*)\end{pmatrix}\notin T_\epsilon^n\right\}\right\}$$
$$=\Pr\{K=0\}$$
$$=\Pr\{K-E[K]=-E[K]\}$$
$$=\Pr\{|K-E[K]|=|-E[K]|\}$$
$$\leq \Pr\{|K-E[K]|\geq |E[K]|\}$$
$$\leq \frac{\text{Var}[K]}{(E[K])^2} \quad \text{(B6)}$$

where the last step follows by the Chebyshev Inequality: $\Pr\{|Y-E[Y]|\geq c\}\leq \frac{\text{Var}[Y]}{c^2}$, where Y is an arbitrary random variable, and c is a constant. Further, we bound $E[K]$ and $\text{Var}[K]$:

$$E[K]=E\left[\sum_{k'=1}^{2^{nR'_{1B}}}\sum_{l'=1}^{2^{nR'_{22}}}I(k',l')\right]$$
$$=\sum_{k'=1}^{2^{nR'_{1B}}}\sum_{l'=1}^{2^{nR'_{22}}}E[I(k',l')]$$
$$=\sum_{k'=1}^{2^{nR'_{1B}}}\sum_{l'=1}^{2^{nR'_{22}}}\Pr\{I(k',l')=1\}$$
$$=\sum_{k'=1}^{2^{nR'_{1B}}}\sum_{l'=1}^{2^{nR'_{22}}}\Pr\left\{\begin{pmatrix}u_{1p}^n(1),u_1^n(1|1),v_1^n(1|1,1,1),\\u_{2p}^n(1|1),u_2^n(1,m^*|1,1,1),\\v_{12}^n(1,k'|1,1,1,1,1,m^*),\\v_2^n(1,l'|1,1,1,1,m^*)\end{pmatrix}\in T_\epsilon^n\right\}$$
$$=\sum_{k'=1}^{2^{nR'_{1B}}}\sum_{l'=1}^{2^{nR'_{22}}}\left(\sum_{(v_{12}^n,v_2^n)\in T_\epsilon^n}p(v_{12}^n|u_{1p}^n,u_1^n,v_1^n,u_{2p}^n,u_2^n)\right.$$
$$\left.\times p(v_2^n|u_{1p}^n,u_1^n,u_{2p}^n,u_2^n)\right)$$

$$\geq \sum_{k'=1}^{2^{nR'_{1B}}}\sum_{l'=1}^{2^{nR'_{22}}}\left(2^{n(H(V_{12},V_2|U_{1p},U_1,V_1,U_{2p},U_2)-\epsilon)}\right.$$
$$\left.\times 2^{-n(H(V_{12}|U_{1p},U_1,V_1,U_{2p},U_2)+\epsilon)}2^{-n(H(V_2|U_{1p},U_1,U_{2p},U_2)+\epsilon)}\right)$$
$$=\sum_{k'=1}^{2^{nR'_{1B}}}\sum_{l'=1}^{2^{nR'_{22}}}\left(\left(2^{-n(I(V_1,V_{12};V_2|U_{1p},U_1,U_{2p},U_2)+3\epsilon)}\right)\right)$$
$$=2^{n(R'_{1B}+R'_{22})}\times\left(2^{-n(I(V_1,V_{12};V_2|U_{1p},U_1,U_{2p},U_2)+3\epsilon)}\right) \quad \text{(B7)}$$

and

$\text{Var}[K]$
$$=\sum_{k'=1}^{2^{nR'_{1B}}}\sum_{\tilde{k}'=1}^{2^{nR'_{1B}}}\sum_{l'=1}^{2^{nR'_{22}}}\sum_{\tilde{l}'=1}^{2^{nR'_{22}}}E[I(k',l')\times I(\tilde{k}',\tilde{l}')]$$
$$\qquad -E[I(k',l')]\times E[I(\tilde{k}',\tilde{l}')]$$
$$=\sum \Pr\{I(k',l')=1,I(\tilde{k}',\tilde{l}')=1\}$$
$$\qquad -\Pr\{I(k',l')=1\}\times\Pr\{I(\tilde{k}',\tilde{l}')=1\} \quad \text{(B8)}$$

The summation in (B8) can be divided into four cases to evaluate:

- If $\tilde{k}'\neq k'$ and $\tilde{l}'\neq l'$,
  $I(k',l')$ is independent of $I(\tilde{k}',\tilde{l}')$. Henceforth, this term contributes nothing to the summation (B8).

- If $\tilde{k}'=k'$ and $\tilde{l}'\neq l'$,
$$\sum \Pr\{I(k',l')=1,I(\tilde{k}',\tilde{l}')=1\}$$
$$\qquad -\Pr\{I(k',l')=1\}\times\Pr\{I(\tilde{k}',\tilde{l}')=1\}$$
$$\leq \sum \Pr\{I(k',l')=1,I(\tilde{k}',\tilde{l}')=1\}$$
$$=\sum \left(\Pr\{I(k',l')=1\}\times\Pr\{I(k',\tilde{l}')=1|I(k',l')=1\}\right)$$
$$=E[K]\times 2^{nR'_{22}}\times\sum_{v_2^n\in T_\epsilon^n}p(v_2^n|u_{1p}^n,u_1^n,u_{2p}^n,u_2^n)$$
$$\leq E[K]\times 2^{nR'_{22}}\times 2^{-n(I(V_1,V_{12};V_2|U_{1p},U_1,U_{2p},U_2)-2\epsilon)} \quad \text{(B9)}$$

Similarly,

- If $\tilde{k}'\neq k'$ and $\tilde{l}'=l'$,
$$\sum \Pr\{I(k',l')=1,I(\tilde{k}',\tilde{l}')=1\}$$
$$\qquad -\Pr\{I(k',l')=1\}\times\Pr\{I(\tilde{k}',\tilde{l}')=1\}$$
$$\leq E[K]\times 2^{nR'_{1B}}\times 2^{-n(I(V_{12};V_2|U_{1p},U_1,U_{2p},U_2,V_1)-2\epsilon)} \quad \text{(B10)}$$

- If $\tilde{k}'=k'$ and $\tilde{l}'=l'$,
$$\sum \Pr\{I(k',l')=1,I(\tilde{k}',\tilde{l}')=1\}$$
$$\qquad -\Pr\{I(k',l')=1\}\times\Pr\{I(\tilde{k}',\tilde{l}')=1\}$$
$$\leq E[K] \quad \text{(B11)}$$

By combing (B7) − (B11) with (B6),

$$\frac{\text{Var}[K]}{(E[K])^2}\leq \frac{1}{2^{nR'_{1B}}\times 2^{-n(5\epsilon)}}+\frac{1}{2^{nR'_{22}}\times 2^{-n(I(V_1;V_2|U_{1p},U_1,U_{2p},U_2)+5\epsilon)}}$$
$$+\frac{1}{2^{n(R'_{1B}+R'_{22})}\times\left(2^{-n(I(V_1,V_{12};V_2|U_{1p},U_1,U_{2p},U_2)+3\epsilon)}\right)} \quad \text{(B12)}$$

Therefore, if

$$R'_{1B} \geq 0 \quad \text{(B13.A)}$$
$$R'_{22} \geq I(V_1;V_2|U_{1p},U_1,U_{2p},U_2) \quad \text{(B13.B)}$$
$$R'_{1B}+R'_{22} \geq I(V_1,V_{12};V_2|U_{1p},U_1,U_{2p},U_2) \quad \text{(B13.C)}$$

TABLE II
ERROR EVENTS OF DECODER 1

| | $\hat{\jmath}_{p_{D_1}}^{(b)}$ | $\hat{\imath}_{D_1}^{(b)}$ | $\left(\hat{m}_{p_{D_1}}^{(b)}, \hat{m}_{D_1}'^{(b)}\right)$ | $\hat{z}_{p_{D_1}}^{(b)}$ | $\left(\hat{k}_{D_1}^{(b)}, \hat{k}_{D_1}'^{(b)}\right)$ |
|---|---|---|---|---|---|
| $E_1^{\text{Decoder 1}}$ | X | – | – | – | – |
| $E_2^{\text{Decoder 1}}$ | 1 | X | X | – | – |
| $E_3^{\text{Decoder 1}}$ | 1 | X | 1 | X | – |
| $E_4^{\text{Decoder 1}}$ | 1 | X | 1 | 1 | – |
| $E_5^{\text{Decoder 1}}$ | 1 | 1 | X | – | – |
| $E_6^{\text{Decoder 1}}$ | 1 | 1 | 1 | X | X |
| $E_7^{\text{Decoder 1}}$ | 1 | 1 | 1 | X | 1 |
| $E_8^{\text{Decoder 1}}$ | 1 | 1 | 1 | 1 | X |

*In Table II and III, when the header of the column contains two indices, an "X" indicates that at least one of the two indices is wrong. It can be observed that the most restrictive case is when both indices are wrongly decoded.

TABLE III
ERROR EVENTS OF DECODER 2

| | $\hat{\jmath}_{D_2}^{(b)}$ | $\left(\hat{m}_{D_2}^{(b)}, \hat{m}_{D_2}'^{(b)}\right)$ | $\left(\hat{\imath}_{D_2}^{(b)}, \hat{\imath}_{D_2}'^{(b)}\right)$ |
|---|---|---|---|
| $E_1^{\text{Decoder 2}}$ | X | – | – |
| $E_2^{\text{Decoder 2}}$ | 1 | X | – |
| $E_3^{\text{Decoder 2}}$ | 1 | 1 | X |

and $n \to \infty$, with sufficiently high probability, encoder 2 can find at least one such pair $(k', l')$.   ∎

## APPENDIX (C)
## PROBABILITY OF ERROR ANALYSIS

$$P_e \equiv \max\{P_{e_1}, P_{e_2}\}$$
$$\leq \Pr\{\text{The Encoding Fails.}\}$$
$$+ \sum_{k=1,2} \Pr\left\{\begin{matrix}\text{The Decoding Error}\\ \text{of Decoder k.}\end{matrix}\bigg|\text{The Encoding Succeeds.}\right\} \quad (C1)$$

As shown in Appendix (A) and (B), the first term is arbitrarily small if (III.2.B) – (III.2.E) are satisfied, and n goes to infinity. Thus, in this part, we focus on bounding the second term, the probability of decoding error. Furthermore, our decoding is performed in $(B+2)$ blocks. According to the similar approach in [3, Section 9.2], it can be shown that each block can be considered separately by assuming that no error was made in the previous blocks. The overall probability of decoding error will be upper-bounded by $(B+2)$ multiplying the maximum probability of decoding error of $(B+2)$ blocks which can be bounded as

$$\max_{b=1,\ldots,(B+2)} \Pr\left\{\begin{matrix}\text{The Decoding Error}\\ \text{of block b.}\end{matrix}\bigg|\begin{matrix}\text{No error was made}\\ \text{prior to block b.}\end{matrix}\right\}$$
$$\leq \max_{b=1,\ldots,(B+2)} \left(\Pr\left\{\begin{matrix}\text{The Decoding Error of}\\ \text{Decoder 1 of block b.}\end{matrix}\bigg|\begin{matrix}\text{No error was made}\\ \text{prior to block b.}\end{matrix}\right\}\right.$$
$$\left.+ \Pr\left\{\begin{matrix}\text{The Decoding Error of}\\ \text{Decoder 2 of block b.}\end{matrix}\bigg|\begin{matrix}\text{No error was made}\\ \text{prior to block b.}\end{matrix}\right\}\right) \quad (C2)$$

We bound the two terms separately:

**[The Decoding Error of Decoder 1]**

Consider the decoding process of block b. Since it is assumed that all the message tuples are equiprobable, without loss of generality, $\left(j_p^{(b)}, i_p^{(b)}, m_p^{(b)}, z_p^{(b)}, k_p^{(b)}\right)$ is assumed to be equal to $(1,1,1,1,1)$ in block b. In the first step of the decoding process of decoder 1, if (III.10) is satisfied, and n is sufficiently large, we have a proper initiation for the backward decoding. In the rest of the decoding process, an error is committed if $\left(\hat{\jmath}_{p_{D_1}}^{(b)}, \hat{\imath}_{p_{D_1}}^{(b)}, \hat{m}_{p_{D_1}}^{(b)}, \hat{z}_{p_{D_1}}^{(b)}, \hat{k}_{p_{D_1}}^{(b)}\right) \neq (1,1,1,1,1)$. We define the error events of decoder 1 in Table II. Note that, in Table II and III, an "X" indicates that the decoding result is error, a "1" indicates that the decoding result is correct, and a dash "–" indicates that it does not matter whether the decoding result is correct or not; in this situation, the most restrictive case is when the decoding result is actually wrong.

Hence, the probability of decoding error of decoder 1 of block b can be bounded as

$$\Pr\left\{\begin{matrix}\text{The Decoding Error of}\\ \text{Decoder 1 of block b.}\end{matrix}\bigg|\begin{matrix}\text{No error was made}\\ \text{prior to block b.}\end{matrix}\right\}$$
$$= \Pr\left\{\begin{matrix}\left(\hat{\jmath}_{p_{D_1}}^{(b)} \neq 1\right) \cup \left(\hat{\imath}_{p_{D_1}}^{(b)} \neq 1\right) \cup \\ \left(\hat{m}_{p_{D_1}}^{(b)} \neq 1\right) \cup \left(\hat{z}_{p_{D_1}}^{(b)} \neq 1\right) \cup \\ \left(\hat{k}_{p_{D_1}}^{(b)} \neq 1\right)\end{matrix}\bigg|\begin{matrix}\text{No error was made}\\ \text{prior to block b.}\end{matrix}\right\}$$
$$\leq \sum_{\text{index}=1}^{8} \Pr\left\{E_{\text{index}}^{\text{Decoder 1}}\bigg|\begin{matrix}\text{No error was made}\\ \text{prior to block b.}\end{matrix}\right\} \quad (C3)$$

The eight terms can be bounded separately:

- When the event $E_k^{\text{Decoder 1}}$, $k=1,2,3,5$, occurs, the quantized version of received symbols at destination 2, $\hat{y}_2^n$, is independent of $y_1^n$. This follows from the fact that either the superposed sequences (i.e., $u_{1p}^n$, $u_1^n$, $u_{2p}^n$ and $u_2^n$) by $\hat{y}_2^n$ are wrongly decoded, or $\hat{z}_{p_{D_1}}^{(b)}$ is wrongly decoded.

  Therefore, the transmitted and received sequences $\left(u_{1p}^n, u_1^n, v_1^n, u_{2p}^n, u_2^n, v_{12}^n, x_3^n, y_1^{n^{(b)}}, \hat{y}_2^n\right)$ are generated iid according to

$$p(u_{1p})p(u_1|u_{1p})p(v_1|u_{1p},u_1)p(u_{2p}|u_{1p})$$
$$\times p(u_2|u_{1p},u_1,u_{2p})p(v_{12}|u_{1p},u_1,v_1,u_{2p},u_2) \quad (C4)$$
$$\times p(x_3|u_{1p},u_{2p})p(y_1|\star)p(\hat{y}_2|u_{1p},u_1,u_{2p},u_2,x_3)$$

  where "$\star$" represents the random variables associated with the sub-messages which are decoded correctly. However, the sequences considered at decoder 1 look as if they were generated iid according to

$$p(u_{1p})p(u_1|u_{1p})p(v_1|u_{1p},u_1)p(u_{2p}|u_{1p})$$
$$\times p(u_2|u_{1p},u_1,v_1,u_{2p})p(v_{12}|u_{1p},u_1,v_1,u_{2p},u_2)$$
$$\times p(x_3|u_{1p},u_{2p})p(y_1|u_{1p},u_1,v_1,u_{2p},u_2,v_{12},x_3) \quad (C5)$$
$$\times p(\hat{y}_2|u_{1p},u_1,v_1,u_{2p},u_2,v_{12},x_3,y_1)$$

  Hence,

$$\Pr\{E_1^{\text{Decoder 1}}|\text{No error was made prior to block b.}\}$$
$$\leq 2^{n(R_{1p}+R_{11}+L_{2P}+\hat{R}+L_{1B})}$$
$$\times 2^{-n\left(E\left[\log\frac{p(u_2|u_{1p},u_1,u_{2p},v_1)}{p(u_2|u_{1p},u_1,u_{2p})}\right]-\varepsilon\right)}$$
$$\times 2^{-n\left(E\left[\log\frac{p(y_1|u_{1p},u_1,v_1,u_{2p},u_2,v_{12},x_3)}{p(y_1)}\right]-\varepsilon\right)}$$
$$\times 2^{-n\left(E\left[\log\frac{p(\hat{y}_2|u_{1p},u_1,v_1,u_{2p},u_2,v_{12},x_3,y_1)}{p(\hat{y}_2|u_{1p},u_1,u_{2p},u_2,x_3)}\right]-\varepsilon\right)}$$
$$= 2^{n(R_{1p}+R_{11}+L_{2P}+\hat{R}+L_{1B})} 2^{-n(I(V_1;U_2|U_{1p},U_1,U_{2p})-3\varepsilon)}$$

$$\times 2^{-n\left(I(Y_1;U_{1p},U_1,V_1,U_{2p},U_2,V_{12},X_3)\right)}$$
$$\times 2^{-n\left(I(Y_1,V_1,V_{12};\hat{Y}_2|U_{1p},U_1,U_{2p},U_2,X_3)\right)} \quad (C6)$$

The other terms can be bounded in exactly the same way, so the details are omitted here.

- When $E_4^{\text{Decoder 1}}$ occurs, $\hat{y}_2^n$ is correlated with $y_1^{n(b)}$. Therefore, the transmitted and received sequences are generated iid according to

$$p(u_{1p})p(u_1|u_{1p})p(v_1|u_{1p},u_1)p(u_{2p}|u_{1p})$$
$$\times p(u_2|u_{1p},u_1,u_{2p})p(v_{12}|u_{1p},u_1,v_1,u_{2p},u_2)$$
$$\times p(x_3|u_{1p},u_{2p})p(y_1|u_{1p},u_1,u_{2p},u_2,x_3) \quad (C7)$$
$$\times p(\hat{y}_2|u_{1p},u_1,u_{2p},u_2,x_3,y_1)$$

However, the sequences considered at decoder 1 look as if they were generated iid according to

$$p(u_{1p})p(u_1|u_{1p})p(v_1|u_{1p},u_1)p(u_{2p}|u_{1p})$$
$$\times p(u_2|u_{1p},u_1,u_{2p},v_1)p(v_{12}|u_{1p},u_1,v_1,u_{2p},u_2)$$
$$\times p(x_3|u_{1p},u_{2p})p(y_1|u_{1p},u_1,v_1,u_{2p},u_2,v_{12},x_3) \quad (C8)$$
$$\times p(\hat{y}_2|u_{1p},u_1,v_1,u_{2p},u_2,v_{12},x_3,y_1)$$

Hence,

$$\Pr\{E_4^{\text{Decoder 1}}|\text{No error was made prior to block } b.\}$$
$$\leq 2^{n(R_{11}+L_{1B})}2^{-n(I(U_2;V_1|U_{1p},U_1,U_{2p})-2\varepsilon)}$$
$$\times 2^{-n\left(I(Y_1;V_1,V_{12}|U_{1p},U_1,U_{2p},U_2,X_3)\right)}$$
$$\times 2^{-n\left(I(\hat{Y}_2;V_1,V_{12}|U_{1p},U_1,U_{2p},U_2,X_3,Y_1)\right)} \quad (C9)$$

- When the event $E_k^{\text{Decoder 1}}$, $k = 6,7,8$, occurs, the indices, $\hat{j}_{p_{D_1}}^{(b)}, \hat{i}_{p_{D_1}}^{(b)}, \hat{m}_{p_{D_1}}^{(b)}$ and $\hat{m}_{D_1}^{\prime(b)}$, are correctly decoded simultaneously. Thus, the sequences $(u_{1p}^n, u_1^n, v_1^n, u_{2p}^n, u_2^n)$ pass the Gel'fand-Pinsker binning, so they are jointly typical. In addition, for the events $E_6^{\text{Decoder 1}}$ and $E_7^{\text{Decoder 1}}$, $\hat{y}_2^n$ is independent of $y_1^{n(b)}$ while, for $E_8^{\text{Decoder 1}}$, $\hat{y}_2^n$ and $y_1^{n(b)}$ are correlated together. The probabilities of these three events can be upper-bounded in the same manner as the above, so we omit the detailed proof.

**[The Decoding Error of Decoder 2]**

Consider the decoding process of block b, $b = 1,2,...,B+1$. Since it is assumed that all the submessage tuples are equiprobable, without loss of generality, we assume that $(j^{(b)}, m^{(b)}, l^{(b)}) = (1,1,1)$ in block b. An error is committed if $(\hat{j}_{D2}^{(b)}, \hat{m}_{D2}^{(b)}, \hat{l}_{D2}^{(b)}) \neq (1,1,1)$. We define the error events of decoder 2 in Table III. The probability of decoding error of decoder 2 of block b can be bounded as:

$$\Pr\left\{\begin{array}{l}\text{The Decoding error of}\\ \text{decoder 2 of block } b.\end{array}\middle|\begin{array}{l}\text{No error was made}\\ \text{prior to block } b.\end{array}\right\}$$
$$= \Pr\left\{\begin{array}{l}(\hat{j}_{D2}^{(b)} \neq 1) \cup (\hat{m}_{D2}^{(b)} \neq 1)\\ \cup (\hat{l}_{D2}^{(b)} \neq 1)\end{array}\middle|\begin{array}{l}\text{No error was made}\\ \text{prior to block } b.\end{array}\right\}$$
$$\leq \sum_{\text{index}=1}^{3} \Pr\left\{E_{\text{index}}^{\text{Decoder 2}}\middle|\begin{array}{l}\text{No error was made}\\ \text{prior to block } b.\end{array}\right\} \quad (C10)$$

When any one event occurs, the transmitted and received sequences $(u_{1p}^n, u_1^n, u_{2p}^n, u_2^n, v_2^n, x_3^n, y_2^{n(b)})$ are generated iid according to

$$p(u_{1p})p(u_1|u_{1p})p(u_{2p}|u_{1p})p(u_2,v_2|u_{1p},u_1,u_{2p})$$
$$\times p(x_3|u_{1p},u_{2p})p(y_2|u_{1p},u_{2p},x_3,\star) \quad (C11)$$

However, the sequences considered at decoder 2 look as if they were generated iid according to

$$p(u_{1p})p(u_1|u_{1p})p(u_{2p}|u_{1p})p(u_2,v_2|u_{1p},u_1,u_{2p})$$
$$\times p(x_3|u_{1p},u_{2p})p(y_2|u_{1p},u_1,u_{2p},u_2,v_2,x_3) \quad (C12)$$

Hence,

$$\Pr\{E_1^{\text{Decoder 2}}|\text{No error was made prior to block } b.\}$$
$$\leq 2^{n(R_{1P}+L_{2P}+L_{22})}$$
$$\times 2^{n\left(-E\left[\log\frac{p(y_2|u_{1p},u_1,u_{2p},u_2,v_2,x_3)}{p(y_2|u_{1p},u_{2p},x_3)}\right]+\varepsilon\right)}$$
$$= 2^{n(R_{1P}+L_{2P}+L_{22})}$$
$$\times 2^{-n(I(Y_2;U_1,U_2,U_{22},V_2|U_{1p},U_{2p},X_3)-\varepsilon)} \quad (C13)$$

and the other terms can be upper-bounded similarly.

Lastly, to sum up, if (III.6), (III.8.A) − (III.8.C), (III.10), (III.13.A) − (III.13.H) and (III.15.A) − (III.15.C) are satisfied, and n goes to infinity, the average probability of error, $P_e$, will go to zero. Thus, the achievability is proved. ∎

## APPENDIX (D)
### PROOF OF THEOREM 2

Consider a code $(|W_1|, |W_2|, n, P_e)$ with average probability of error $P_e \to 0$. The probability distribution on the joint ensemble space $W_1 \times W_2 \times \mathcal{X}_1^n \times \mathcal{X}_2^n \times \mathcal{X}_3^n \times \mathcal{Y}_1^n \times \mathcal{Y}_2^n$ is given by

$$p(w_1, w_2, x_1^n, x_2^n, x_3^n, y_1^n, y_2^n)$$
$$= p(w_1)\, p(w_2)\, p(x_1^n|w_1)p(x_2^n|w_1, w_2)$$
$$\times \prod_{i=1}^{n}\left(I\left(x_{3_i} = f_{3_i}(y_2^{i-1})\right) \times p(y_{1_i}, y_{2_i}|x_{1_i}, x_{2_i}, x_{3_i})\right) \quad (D1)$$

where $I(\cdot)$ is the indicator function which equals to 1 if its argument is true and equals to 0 otherwise. Additionally, $p(x_1^n|w_1)$ and $p(x_2^n|w_1, w_2)$ are either 0 or 1.

According to Fano's inequality [2, theorem 2.10.1], we have

$$H(W_1|Y_1^n) \leq nR_1P_e + 1 \triangleq n\varepsilon_1 \quad (D2)$$
$$H(W_2|Y_2^n) \leq nR_2P_e + 1 \triangleq n\varepsilon_2 \quad (D3)$$

where $\varepsilon_1$ and $\varepsilon_2$ goes to zero if $P_e \to 0$. We can now bound the rate $R_1$ as

$$nR_1 = H(W_1)$$
$$\leq I(W_1; Y_1^n) + H(W_1|Y_1^n)$$
$$= \sum_{i=1}^{n} I(Y_{1_i}; W_1|Y_1^{i-1}) + n\varepsilon_1$$
$$= \sum_{i=1}^{n} H(Y_{1_i}|Y_1^{i-1}) - H(Y_{1_i}|W_1, Y_1^{i-1}) + n\varepsilon_1$$
$$\leq \sum_{i=1}^{n} H(Y_{1_i}) - H(Y_{1_i}|W_1, Y_1^{i-1}, X_{1_i}, X_{2_i}, X_{3_i}) + n\varepsilon_1 \quad (D4)$$
$$= \sum_{i=1}^{n} I(Y_{1_i}; X_{1_i}, X_{2_i}, X_{3_i}) + n\varepsilon_1 \quad (D5)$$

where (D4) follows since conditioning does not increase entropy, and (D5) follows from the Markov chain $(W_1, Y_1^{i-1}) - (X_{1_i}, X_{2_i}, X_{3_i}) - Y_{2_i}$.

We can similarly derive the upper bound for $R_2$ as

$nR_2 = H(W_2)$
$= I(W_2; Y_2^n) + H(W_2|Y_2^n)$
$\leq I(W_1, Y_2^n; W_2) + n\varepsilon_2$
$= I(Y_2^n; W_2|W_1) + n\varepsilon_2$ \hfill (D6)
$= I\big(Y_2^n; W_2, X_2^n(W_1, W_2)\big|W_1, X_1^n(W_1)\big) + n\varepsilon_2$
$= \sum_{i=1}^{n} I\big(Y_{2_i}; W_2, X_2^n\big|W_1, X_1^n, Y_2^{i-1}, X_{3_i}\big) + n\varepsilon_2$ \hfill (D7)
$= \sum_{i=1}^{n} \Big(H\big(Y_{2_i}|W_1, X_1^n, Y_2^{i-1}, X_{3_i}\big) - H\big(Y_{2_i}|W_1, W_2, X_1^n, X_2^n, X_{3_i}, Y_2^{i-1}\big)\Big)$
$\quad + n\varepsilon_2$
$\leq \sum_{i=1}^{n} H\big(Y_{2_i}|X_{1_i}, X_{3_i}\big) - H\big(Y_{2_i}|X_{1_i}, X_{2_i}, X_{3_i}\big) + n\varepsilon_2$ \hfill (D8)
$\leq \sum_{i=1}^{n} I\big(Y_{2_i}; X_{2_i}|X_{1_i}, X_{3_i}\big) + n\varepsilon_2$ \hfill (D9)

where (D6) follows from the independence between $W_1$ and $W_2$, (D7) follows from $x_{3_i} = f_{3_i}(y_2^{i-1})$, and (D8) follows from the Markov chain $\big(W_1, W_2, X_1^{i-1}, X_2^{i-1}, Y_2^{i-1}\big) - \big(X_{1_i}, X_{2_i}, X_{3_i}\big) - Y_{2_i}$.

We next derive an upper bound on the sum rate $R_1 + R_2$

$nR_1 + nR_2 = H(W_1, W_2)$
$= I(W_1, W_2; Y_1^n) + H(W_1, W_2|Y_1^n) + I(W_1, W_2; Y_2^n|Y_1^n)$
$\quad - I(W_1, W_2; Y_2^n|Y_1^n)$
$= I(W_1, W_2; Y_1^n, Y_2^n) + H(W_1, W_2|Y_1^n) - I(W_1, W_2; Y_2^n|Y_1^n)$
$= I(W_1, W_2; Y_1^n, Y_2^n) + H(W_1, W_2|Y_1^n, Y_2^n)$
$= I(W_1, W_2; Y_1^n, Y_2^n) + H(W_1|Y_1^n, Y_2^n) + H(W_2|W_1, Y_1^n, Y_2^n)$
$\leq I(W_1, W_2; Y_1^n, Y_2^n) + H(W_1|Y_1^n) + H(W_2|Y_2^n)$
$\leq I(W_1, W_2; Y_1^n, Y_2^n) + n(\varepsilon_1 + \varepsilon_2)$
$= \sum_{i=1}^{n} I\big(Y_{1_i}, Y_{2_i}; W_1, W_2|Y_1^{i-1}, Y_2^{i-1}, X_{3_i}\big) + n(\varepsilon_1 + \varepsilon_2)$ \hfill (D10)
$\leq \sum_{i=1}^{n} H\big(Y_{1_i}, Y_{2_i}|X_{3_i}\big)$
$\quad - H\big(Y_{1_i}, Y_{2_i}|W_1, W_2, Y_1^{i-1}, Y_2^{i-1}, X_{1_i}, X_{2_i}, X_{3_i}\big)$ \hfill (D11)
$= \sum_{i=1}^{n} H\big(Y_{1_i}, Y_{2_i}|X_{3_i}\big) - H\big(Y_{1_i}, Y_{2_i}|X_{1_i}, X_{2_i}, X_{3_i}\big)$ \hfill (D12)
$= \sum_{i=1}^{n} I\big(Y_{1_i}, Y_{2_i}; X_{1_i}, X_{2_i}|X_{3_i}\big)$ \hfill (D13)

where (D10) follows from $x_{3_i} = f_{3_i}(y_2^{i-1})$, (D11) follows since conditioning does not increase entropy, and (D12) follows from the Markov chain $\big(W_1, W_2, Y_1^{i-1}, Y_2^{i-1}\big) - \big(X_{1_i}, X_{2_i}, X_{3_i}\big) - \big(Y_{1_i}, Y_{2_i}\big)$.

At last, the upper bounds (D5), (D9) and (D13) can be transformed into the single-letter bounds in theorem 2 by introducing an auxiliary random variable Q which is independent of $(w_1, w_2, x_1^n, x_2^n, x_3^n, y_1^n, y_2^n)$ and uniformly distributed over $\{1, 2, \ldots, n\}$. Define $X_1 \triangleq X_{1Q}, X_2 \triangleq X_{2Q}, X_3 \triangleq X_{3Q}, Y_1 \triangleq Y_{1Q}$ and $Y_2 \triangleq Y_{2Q}$. Then for (D5)

$\frac{1}{n} \sum_{i=1}^{n} I\big(Y_{1_i}; X_{1_i}, X_{2_i}, X_{3_i}\big) = I(Y_1; X_1, X_2, X_3|Q)$
$\leq I(Y_1; X_1, X_2, X_3)$ \hfill (D14)

by the Markov chain

$$Q - (X_1, X_2, X_3) - Y_1 \hfill (D15)$$

The other bounds can be transformed in exactly the same manner, and theorem 2 is proved. ∎